\documentclass{article}

\usepackage{microtype}
\usepackage{graphicx}
\usepackage{subcaption}
\usepackage{booktabs} 

\usepackage{hyperref}

\usepackage[preprint]{icml2026}

\usepackage{amsmath}
\usepackage{amssymb}
\usepackage{mathtools}
\usepackage{amsthm}

\usepackage{adjustbox}
\usepackage{algorithm}
\usepackage{algorithmic}
\usepackage{booktabs}   
\usepackage{array}      
\usepackage{graphicx}   
\usepackage{subcaption}

\usepackage[capitalize,noabbrev]{cleveref}

\theoremstyle{plain}

\theoremstyle{definition}
\newtheorem{definition}{Definition}

\theoremstyle{remark}

\usepackage[textsize=tiny]{todonotes}

\icmltitlerunning{Alpha Discovery via Grammar-Guided Learning and Search}

\begin{document}

\twocolumn[
  \icmltitle{Alpha Discovery via Grammar-Guided Learning and Search}



  \icmlsetsymbol{equal}{*}

  \begin{icmlauthorlist}
    \icmlauthor{Han Yang}{yyy}
    \icmlauthor{Dong Hao}{yyy}
    \icmlauthor{Zhuohan Wang}{sch}
    \icmlauthor{Qi Shi}{comp}
    \icmlauthor{Xingtong Li}{yyy}
  \end{icmlauthorlist}

  \icmlaffiliation{yyy}{University of Electronic Science and Technology of China, Chengdu, China}
  \icmlaffiliation{sch}{King's College London, London, United Kingdom}
  \icmlaffiliation{comp}{University of Southampton, Southampton, United Kingdom}

  \icmlcorrespondingauthor{Han Yang}{yanghan667@std.uestc.edu.cn}
  \icmlcorrespondingauthor{Dong Hao}{haodong@uestc.edu.cn}
  \icmlcorrespondingauthor{Zhuohan Wang}{zhuohan.wang@kcl.ac.uk}
  \icmlcorrespondingauthor{Qi Shi}{qi.shi@soton.ac.uk}
  \icmlcorrespondingauthor{Xingtong Li}{lixingtong@std.uestc.edu.cn}

  \icmlkeywords{Machine Learning, ICML}

  \vskip 0.3in
]



\printAffiliationsAndNotice{}  

\begin{abstract}
Automatically discovering formulaic alpha factors is a central problem in quantitative finance. Existing methods often ignore syntactic and semantic constraints, relying on exhaustive search over unstructured and unbounded spaces. We present AlphaCFG, a grammar-based framework for defining and discovering alpha factors that are syntactically valid, financially interpretable, and computationally efficient. AlphaCFG uses an alpha-oriented context-free grammar to define a tree-structured, size-controlled search space, and formulates alpha discovery as a tree-structured linguistic Markov decision process, which is then solved using a grammar-aware Monte Carlo Tree Search guided by syntax-sensitive value and policy networks. Experiments on Chinese and U.S. stock market datasets show that AlphaCFG outperforms state-of-the-art baselines in both search efficiency and trading profitability. Beyond trading strategies, AlphaCFG serves as a general framework for symbolic factor discovery and refinement across quantitative finance, including asset pricing and portfolio construction.

\end{abstract}

\section{Introduction}

\subsection{Alpha discovery}

In quantitative finance, alpha factors play a central role in asset management, quantitative trading, and investment decision-making. An alpha factor is an explicit function that maps historical market features, such as prices and volumes, to predictions of future returns. Alpha discovery refers to the systematic identification of such predictive functions from historical data and remains a core challenge due to the vast and complex space of possible functional forms. Beyond their practical importance, alpha discovery poses a fundamental machine-learning challenge: identifying symbolic functions that are both predictive and interpretable under severe combinatorial constraints.

Existing approaches to alpha discovery can be broadly classified into three categories. \textit{Heuristic or expert-driven methods} rely on financial intuition, such as value factors (e.g., price-to-earnings ratios \citep{fama1992cross}) and momentum factors (e.g., past 12-month returns \citep{carhart1997persistence}), but lack scalability and are quickly arbitraged once widely adopted, reducing  predictive accuracy over time. \textit{Data-driven learning methods}, including regression \citep{bhandari2022predicting,qin2017dual,dai2022price,mozaffari2024predictive}, tree-based ensembles \citep{wang2023xgboost,bisdoulis2024assets}, unsupervised learning \citep{xu2025unsupervised}, and reinforcement learning \citep{lee2001stock}, can capture complex nonlinear patterns, yet often suffer from limited interpretability and overfitting due to their black-box nature. \textit{Formulaic alpha methods} \citep{zhang2020autoalpha} emphasize human-readable mathematical expressions composed of predefined operators, offering transparency and interpretability, and have therefore regained recent attention.


\vspace{-4pt}
\begin{table}[!htbp]
\centering
\caption{Comparison of Alpha Discovery Methods}
\resizebox{\linewidth}{!}{%
\begin{tabular}{lll}
\toprule
\textbf{Category} & \textbf{Pros} & \textbf{Cons} \\
\midrule
Heuristic / Expert & Intuitive, easy to use & Limited, quickly arbitraged \\
Data-driven Learning & Captures complex patterns & Black-box, less interpretable \\
Formulaic Alpha & Interpretable, transparent & Computationally expensive \\
\bottomrule
\end{tabular}
}\vspace{-8pt}
\end{table}

Our work lies at the intersection of data-driven learning and formulaic alpha methods, aiming at the \textit{automatic discovery of explainable alpha factors}. This problem can be viewed as symbolic regression \citep{makke2024interpretable}, which seeks \textit{explicit mathematical expressions that fit data while remaining interpretable},  but is  difficult due to its combinatorial search space and semantic equivalence among expressions. Early approaches such as genetic programming (GP) \citep{zhang2020autoalpha} evolve expression trees to optimize information coefficients. More recent methods, including AlphaGen \citep{yu2023generating} and AlphaQCM \citep{zhu2025alphaqcm}, adopt reinforcement learning to improve scalability.
Existing methods  face  the following fundamental challenges.

(1) \textit{Lack of linguistic characterization leads to inefficient search in an unbounded space.}
Automated discovery of formulaic alphas is fundamentally a problem of searching over mathematical languages, yet existing methods lack an explicit linguistic framework to organize and constrain this search. In the absence of formal grammatical structure, current approaches must explore vast, and often effectively infinite, combinatorial spaces of expressions, relying on ad hoc syntactic checks to ensure validity. This unstructured exploration severely limits sample efficiency, degrades model performance, and incurs substantial computational cost.

(2) \textit{Semantic redundancy causes systematic waste in learning and search.}
Many syntactically distinct mathematical sequences correspond to the same underlying semantics, but existing methods mostly encode expressions as linear sequences and treat the  variants as independent. As a result, semantically equivalent expressions are repeatedly explored and evaluated, leading to significant redundancy in representation learning and search, and greatly reducing  efficiency.

\subsection{Our Work}

We propose AlphaCFG,\footnote{\sloppy Our source code is available at \url{https://github.com/HanYang544/AlphaCFG}} a general linguistic–learning framework for the automatic discovery of interpretable alpha factors. The central idea is to treat alpha discovery as a structured language generation and learning problem, rather than an unstructured search over mathematical expressions. By combining formal grammar with learning and search, AlphaCFG provides a principled way to generate, validate, and optimize human-readable alpha factors.  In this way, grammar serves as an explicit inductive bias that shapes both the search space and the learning dynamics.

(1) \textit{\textbf{Grammar-Constrained Alpha Factors.}}
From a language-theoretic perspective, we first formalize the space of alpha factors as a structured mathematical language.
We propose two formal languages, $\alpha$-Syn and $\alpha$-Sem, that integrate context-free grammar (CFG) with finance domain–specific knowledge of alpha factors. $\alpha$-Syn enforces grammatical correctness, while $\alpha$-Sem further ensures financial semantic validity. These languages generate alpha expressions recursively in a tree-structured form, making tree-structure–based learning and optimization possible. To control complexity and reduce redundancy, we further enforce (i) length constraints to bound the search space, and (ii) expression-tree pruning to remove syntactically distinct but semantically equivalent factors.

(2) \textit{\textbf{Structure  Characterization of Alpha Space}}.
Building on this grammar-based language, we cast alpha discovery as a large Tree-Structured Linguistic Markov Decision Process (TSL-MDP), where each state is a partial expression, terminal states represent complete alpha factors, and rewards are given by the information coefficient (IC) on real market data. This formulation transforms alpha discovery from unstructured trial-and-error into a principled sequential decision process over the space of formulaic alpha factors.

(3) \textit{\textbf{Reinforcing MCTS with Syntax-Aware Learning}}.
Finally, we design a learning and search algorithm that exploits the grammar-induced structure of the TSL-MDP. We employ a grammar-aware Monte Carlo Tree Search (MCTS), in which action selection is guided by a syntax-aware Upper Confidence Bound (UCB) rule. To generalize across the large state space, each partial expression tree is encoded using a Tree-LSTM, yielding structure-aware representations shared by a value network, which estimates expected performance from historical  market data, and a policy network, which predicts promising alpha expansions. Through reinforced interaction between MCTS and these learned models, AlphaCFG progressively refines its search strategy and discovers high-quality alpha factors efficiently.
 
AlphaCFG is not limited to trading strategies and naturally extends to other quantitative finance tasks, by allowing flexible customization of operators, grammatical structures, and objective functions. We use trading as a representative  testbed to demonstrate the effectiveness of AlphaCFG.
We evaluate AlphaCFG on CSI~300 and S\&P~500 stocks, where it consistently outperforms strong baselines across multiple metrics, including returns, information coefficient (IC), Sharpe ratio, and maximum drawdown. Our results show that improved grammar design yields faster convergence and higher-quality factors. Moreover, AlphaCFG \textit{effectively refines existing factors and improves their predictive performance}, highlighting its utility as a general tool for factor refinement and augmentation. Ablation studies further confirm the critical roles of grammar design and syntax-based representation learning in effective factor discovery.

\section{Problem Formulation}

Consider a market with $n$ stocks  over $T$ trading days. 
On each day $t \in \{1,2,\dots,T\}$, stock $i$ is associated with a feature matrix 
$\mathbf{x}_{t,i} \in \mathbb{R}^{m \times \tau'}$, which records $m$ raw market features 
(e.g., opening and closing prices, volumes) over $\tau'$ days.
We denote by 
$\mathbf{X}_t = (\mathbf{x}_{t,1}, \mathbf{x}_{t,2}, \dots, \mathbf{x}_{t,n})$
the collection of features for all stocks on day $t$.

An \emph{alpha factor} is a function
$f: \mathbb{R}^{m \times \tau'} \rightarrow \mathbb{R}$
that maps the historical features of a single stock to a scalar score.
Applying $f$ cross-sectionally to all stocks on day $t$ yields a factor vector
$\mathbf{y}_t = (y_{t,1}, \dots, y_{t,n}) \in \mathbb{R}^n$,
where $y_{t,i} = f(\mathbf{x}_{t,i})$ (illustrated in \Cref{fig:alpha_factor}).
These scores are subsequently used to rank stocks or construct portfolios.

We focus on \emph{formulaic} alpha factors, which are explicit mathematical expressions constructed from a predefined set of input features (\Cref{tab:stock_features}), constants (\Cref{tab:constants}), and operators (\Cref{tab:operators}).
These operators and operands are commonly used in quantitative finance \citep{yang2020qlib}.
Representative examples are shown in \Cref{fig:example}.


\paragraph{Evaluation via Information Coefficient.}
The predictive quality of an alpha factor is evaluated using the \emph{Information Coefficient} (IC), a standard metric in asset management \citep{grinold2000active}.
For a given prediction horizon $\tau$, the realized $\tau$-day return of stock $i$ observed at day $t$ is
\begin{equation}
r_{t,i}^{(\tau)} = \frac{\mathrm{Close}_{t+\tau,i}}{\mathrm{Close}_{t,i}} - 1,
\end{equation}
where $\mathrm{Close}_{t,i}$ is the closing price of stock $i$ on day $t$.
Let $\mathbf{r}_t^{(\tau)} = (r_{t,1}^{(\tau)}, \dots, r_{t,n}^{(\tau)})$ denote the cross-sectional return vector.
The daily IC at day $t$ is defined as the Pearson correlation between factor scores and subsequent returns:
\begin{equation}
\resizebox{\hsize}{!}{$
\mathrm{IC}_{t}(\mathbf{y}_t,\mathbf{r}_t^{(\tau)})
= \frac{\sum_{i=1}^n (y_{t,i}-\bar{y}_t)(r_{t,i}^{(\tau)}-\overline{r}_t^{(\tau)})}
{\sqrt{\sum_{i=1}^n (y_{t,i}-\bar{y}_t)^2}
 \sqrt{\sum_{i=1}^n (r_{t,i}^{(\tau)}-\overline{r}_t^{(\tau)})^2}},
$}\nonumber
\label{eq:ic-definition}
\end{equation}
where
$\bar{y}_t = \frac{1}{n}\sum_{i=1}^n y_{t,i}$ and
$\overline{r}_t^{(\tau)} = \frac{1}{n}\sum_{i=1}^n r_{t,i}^{(\tau)}$.


To assess factor performance over the entire period, we use average daily IC of alpha factor $f$:
\begin{equation}\label{eq_icf}
\mathrm{IC}(f) = \frac{1}{T}\sum_{t=1}^T \mathrm{IC}_{t}\bigl(\mathbf{y}_t,\mathbf{r}_t^{(\tau)}\bigr).
\end{equation}
A higher $\mathrm{IC}(f)$ indicates stronger predictive power.
Accordingly, the goal of \emph{alpha discovery} is to identify formulaic factors that maximize $\mathrm{IC}(f)$.


In practice, a common and effective strategy is to linearly combine multiple factors.
Following AlphaGen \citep{yu2023generating}, we optimize the IC of such linear combinations (referred to as a \emph{factor pool}).
The detailed combination procedure is provided in \Cref{alg:incremental-combination} in Appendix~\ref{Appendix-section-Linear-Combination}.


\section{Design Language of Interpretable Alphas}

The space of formulaic alpha factors grows combinatorially with expression length, rendering brute-force  search  inefficient.
Moreover, a large fraction of candidate expressions are either \emph{syntactically invalid} (i.e., ill-formed operator compositions) or \emph{semantically nonsensical} (i.e., violating financial or temporal constraints), which severely hampers both efficiency and interpretability.

From a machine learning perspective, automated alpha discovery is therefore not merely an optimization problem, but fundamentally a \emph{language design problem}: one must define a hypothesis space that is expressive enough to capture meaningful financial signals, while being sufficiently structured to admit efficient search and learning.
In the absence of such structure, existing methods are forced to explore an effectively unbounded symbolic space, leading to severe combinatorial explosion and redundant evaluations.

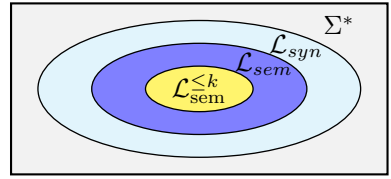
\begin{figure}[htbp]
\centering
\begin{tikzpicture}[scale=0.65, xscale=1.1, yscale=0.7]
    \fill[gray!10] (-3.5, -2.5) rectangle (3.5, 2.5);
    \draw[thick] (-3.5, -2.5) rectangle (3.5, 2.5);
    \node at (2.6, 1.857) [black, font=\normalsize] {$\Sigma^*$};

    \fill[cyan!10] (0,0) ellipse (3cm and 2cm);
    \draw[black, line width=0.5pt] (0,0) ellipse (3cm and 2cm);
    \node at (1.8, 1.214) [text=black, font=\normalsize] {$\mathcal{L}_{syn}$};

    \fill[blue!50!white]
    (0,0) ellipse (2cm and 1.333cm);
    \draw[black, line width=0.5pt] (0,0) ellipse (2cm and 1.333cm);
    \node at (1.2, 0.724) [text=black, font=\normalsize] {$\mathcal{L}_{sem}$};

    \fill[yellow!70]
    (0,0) ellipse (1cm and 0.666cm);
    \draw[black, line width=0.5pt] (0,0) ellipse (1cm and 0.666cm);
    \node at (0, 0) [text=black, font=\normalsize] {$\mathcal{L}_{\mathrm{sem}}^{\le k}$};
\end{tikzpicture}
\caption{
Nested spaces of alpha expressions:
$\Sigma^*$ (all symbol sequences),
$\mathcal{L}_{\mathrm{syn}}$ (syntactically valid),
$\mathcal{L}_{\mathrm{sem}}$ (semantically valid),
and $\mathcal{L}_{\mathrm{sem}}^{\le K}$ (length-bounded semantic alphas).
}
\label{fig:nested_ellipse}
\end{figure}

To address these challenges, we introduce a formal \emph{linguistic characterization} of alpha factors based on \textit{Context-Free Grammar (CFG)} \citep{chomsky1959algebraic,hopcroft1979automata}.
By explicitly specifying the syntactic rules that govern valid alpha expressions, we restrict the search space to well-formed expressions, enforce operator-operand consistency, and enable tree-based search and learning.
This linguistic view allows us to systematically decompose the alpha search space into nested levels of validity, as illustrated in \Cref{fig:nested_ellipse}.

\subsection{Syntactically-Valid Alpha Language} 
We begin by defining a grammar that ensures \emph{syntactic validity}, which serves as the foundation for the following sections of semantic constraints and learning algorithms.

Syntactic validity requires that every generated alpha expression be a well-formed and evaluable symbolic program.
It entails two conditions:
(i) a well-defined hierarchical structure enforced by prefix notation and recursive nonterminal expansion; and
(ii) strictly follow operator arity, so that each operator receives the correct number of operands.
These  are captured by the following generation rule:
\begin{equation}
\label{fl-abstract-generation}
\mathsf{Expr} \;\to\; \mathsf{Op}(\mathsf{Expr}, \dots) \;\mid\; \mathsf{TermSyb},
\end{equation}
where $\mathsf{Expr} \in \mathcal{N}$ denotes a recursively expandable nonterminal symbol,
$\mathsf{Op} \in \mathcal{T}$ denotes prefix-notation operators,
and $\mathsf{TermSyb} \in \mathcal{T}$ denotes terminal symbols which are  features and constants.



\paragraph{Structural Well-Formedness.}
Formula~(\ref{fl-abstract-generation}) enforces a \textit{prefix-notation} structure in which each operator $\mathsf{Op}$ precedes its operands, eliminating ambiguity in operator precedence and evaluation order.
Recursive expansion of $\mathsf{Expr}$ enables the construction of complex expressions, while termination is ensured by substituting terminal symbols.
Therefore, each valid derivation admits a unique hierarchical representation which we call  \emph{Abstract Syntax Representation (ASR)}. \footnote{In formal language \citep{hopcroft1979automata}, an expression corresponds to an abstract syntax tree (AST); we use the term ASR to distinguish it from the  large search tree introduced later.}

\begin{definition}\label{def:expression tree}
An Abstract Syntax Representation (ASR) is a rooted, ordered tree encoding a single alpha expression, whose internal nodes are operators with arity-matched children and whose leaves are features, constants, or (in partial derivations) nonterminal symbols.
\end{definition}

\paragraph{Operator Arity Constraints.}
Syntactic validity also requires that all operators be applied with the correct number of operands.
We instantiate $\mathsf{Op}$ using operator families with fixed arity, reflecting common primitives in quantitative finance.
These include unary operators ($\mathsf{UnaryOp}$), binary operators ($\mathsf{BinaryOp}$),
rolling operators ($\mathsf{RollingOp}$),
paired rolling operators ($\mathsf{PairedRollingOp}$),
and nullary terminal symbols ($\mathsf{TermSyb}$) representing constants and raw features.
The resulting production rules are given by:
\begin{equation}
\label{fl-basic-production-rule}
\resizebox{0.8\hsize}{!}{$
\begin{aligned}
\mathsf{Expr} \;\to\;
&\mathsf{UnaryOp}(\mathsf{Expr})
\;\mid\; \mathsf{BinaryOp}(\mathsf{Expr}, \mathsf{Expr}) \\
&\mid\; \mathsf{RollingOp}(\mathsf{Expr}, \mathsf{Expr}) \\
&\mid\; \mathsf{PairedRollingOp}(\mathsf{Expr}, \mathsf{Expr}, \mathsf{Expr}) \\
&\mid\; \mathsf{TermSyb}.
\end{aligned}
$}
\end{equation}

All feature symbols and constants are listed in \Cref{tab:stock_features} and \Cref{tab:constants}, respectively,
while \Cref{tab:operators} lists all operators together with their  arity categories.
These rules fully specify the admissible syntactic forms of alpha expressions.


We now formally define the context-free grammar that characterizes syntactically valid alpha expressions.


\begin{definition}[$\alpha$-Syn]\label{df-alpha-CFG}
 The context-free grammar for a syntactically valid alpha language $\alpha$-Syn is defined as $G=(\mathcal{N},\mathcal{T},\mathcal{P},\mathcal{S})$ where
$\mathcal{N}$ is the recursively expandable nonterminal symbols,
$\mathcal{T}$ is the terminal symbols including stock features (\Cref{tab:stock_features}),
numerical constants (\Cref{tab:constants}), and operators with fixed arity (\Cref{tab:operators}),
$\mathcal{P}$ is the production rules given in Formula \ref{fl-basic-production-rule},
which enforce prefix-notation and strict operator-arity consistency, and  $\mathcal{S}$ is the start symbol.
\end{definition}

\Cref{df-alpha-CFG} generates the language $\mathcal{L}_{\mathrm{syn}}$ of syntactically valid alpha expressions, as illustrated in \Cref{fig:nested_ellipse}.

\subsection{Semantically-Interpretable Alpha Language}
While $\alpha$-Syn guarantees syntactic validity, it does not ensure semantic soundness in quantitative trading, as syntax alone cannot capture domain-specific financial constraints such as temporal coherence, numerical admissibility, or economically meaningful operator interactions.
Now we refine $\alpha$-Syn in \Cref{df-alpha-CFG} by embedding domain-informed semantic constraints directly into the grammar, thereby defining a semantically interpretable alpha language.

\paragraph{Semantic Constraints.}
We enforce a set of minimal and widely accepted financial semantic constraints:
(i) \textit{Rolling window constraint}: the window-size operand of $\mathsf{RollingOp}$ and $\mathsf{PairedRollingOp}$ is integer constant;
(ii) \textit{Non-triviality}: expressions cannot consist solely of constants and operators;
(iii) \textit{Numerical validity}: operands must lie within domains consistent with their operators;
(iv) \textit{Time-series consistency}: $\mathsf{PairedRollingOp}$ must operate on two time-varying features; constant operands are disallowed.

\begin{definition}[$\alpha$-Sem]\label{df:alpha-CFG-semantic}
A semantic refinement of $\alpha$-Syn  is a context-free grammar
$G=(\mathcal{N}',\mathcal{T}',\mathcal{P}',\mathcal{S})$ that shares the same start symbol $\mathcal{S}$ as $\alpha$-Syn. The nonterminal symbols $\mathcal{T}'$ add Num and Constant.
The Terminal symbols $\mathcal{T}$ is refined to containing  features in \Cref{tab:stock_features}, constant in \Cref{tab:constants} and  operators in \Cref{tab:operators}.
The production rules $\mathcal{P}'$ distinguishes the type of operands by: 
\begin{equation}
\label{fl-semantic-production-rule}
\resizebox{1\hsize}{!}{$
\begin{aligned}
\mathsf{Expr} \;\to\;&\; \mathsf{Feature}
\;\mid\; \mathsf{UnaryOp}(\mathsf{Expr}) \\
&\mid\; \mathsf{BinaryOp}(\mathsf{Expr}, \mathsf{Expr})
\mid \mathsf{BinaryOp}(\mathsf{Expr}, \mathsf{Constant}) \\
&\mid\; \mathsf{BinaryOp\_Asym}(\mathsf{Constant}, \mathsf{Expr}) \\
&\mid\; \mathsf{RollingOp}(\mathsf{Expr}, \mathsf{Num})\\
&\mid\; \mathsf{PairedRollingOp}(\mathsf{Expr}, \mathsf{Expr}, \mathsf{Num}), \\
\mathsf{Num} \;\to\;&\; 20 \mid \dots,
\qquad
\mathsf{Constant} \;\to\; -0.01 \mid \dots
\end{aligned}\nonumber
$}
\end{equation}
\end{definition}

The terminal symbols and operators of $\alpha$-Sem can be revised or extended beyond \Cref{tab:stock_features}, \Cref{tab:constants}, and \Cref{tab:operators} to support other domains or tasks in quantitative finance. 
\paragraph{Length Bounded Grammar $\alpha$-Sem-$k$.}
Although $\alpha$-Sem enforces both syntactic and semantic validity, its recursive production rules can still generate expressions of unbounded depth, leading to an intractable search space.
We apply \emph{$k$-bounded constraint}, which assign each alpha expression with a length counter $k$ capped at $K$. Each production rule incurs an incremental cost $\Delta k$ (\Cref{tab:delta-depth}), and a rule may be applied only if $k+\Delta k \le K$.
This constraint yields a bounded semantic grammar $\alpha$-Sem-$k$ (\Cref{alg:k-bounded-derivation}).

\subsection{Alpha Space Structure}
\label{sec:language-level}
Each grammar above ($\alpha$-Syn, $\alpha$-Sem, and $\alpha$-Sem-$k$) generates a space of many alpha expressions, corresponding to a \emph{formal alpha language}, denoted by
$\mathcal{L}_{\mathrm{syn}}$, $\mathcal{L}_{\mathrm{sem}}$, and $\mathcal{L}_{\mathrm{sem}}^{\le K}$, respectively.
These languages are naturally nested, as illustrated in \Cref{fig:nested_ellipse}, with each successive layer imposing additional constraints and yielding a progressively smaller and more structured hypothesis space for alpha discovery.

Among them, the $k$-bounded semantic grammar $\alpha$-Sem-$k$ plays a central role in this work.
Bounding the derivation depth while enforcing both syntactic and semantic validity yields a finite yet expressive language $\mathcal{L}_{\mathrm{sem}}^{\le K}$, enabling systematic search.
A detailed analysis of the space complexity of $\mathcal{L}_{\mathrm{sem}}^{\le K}$ is provided in Appendix~\ref{sec:analysis}.

\begin{definition}[Search Space Structure]
\label{def:language_tree}
Given a grammar $\alpha$-Syn, $\alpha$-Sem, or $\alpha$-Sem-$k$, the corresponding alpha language can be represented as a rooted tree.
The root node corresponds to the start symbol, each edge corresponds to the application of a production rule,
intermediate nodes represent partially derived expressions, and leaf nodes correspond to fully derived alpha factors.
\end{definition}

This formulation of \Cref{def:language_tree} fundamentally changes the nature of alpha discovery.
Rather than exhaustive searching over the unstructured and infinite symbol space $\Sigma^*$, alpha discovery can now be viewed as exploring a tree-structured language space
$\mathcal{L}_{\mathrm{sem}}^{\le K}$ induced by $\alpha$-Sem-$k$.
In this view, our alpha discovery reduces to identifying high-quality leaf nodes within a large but well-organized derivation tree.
\begin{figure}[t]
\centering
\includegraphics[width=1\linewidth]{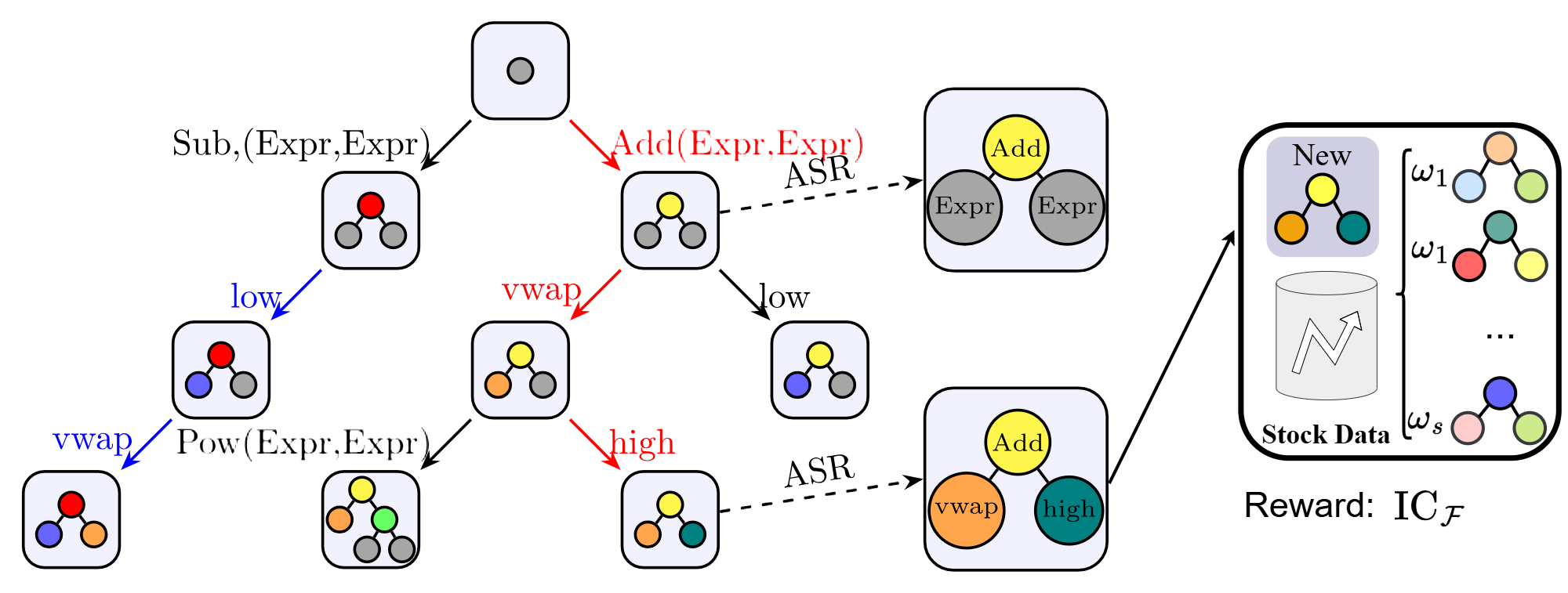}
\caption{The tree-structured search space.}
\label{fig:tree search space}
\end{figure}

\Cref{fig:tree search space} illustrates the structure of the  space
 of all alpha expressions under $\alpha$-Sem-$k$. In this tree, each rounded-box node corresponds to an alpha expression, which is equivalent to an Abstract Syntax Representation (ASR) shown in the middle of the figure. Within each ASR, grey nodes denote nonterminal symbols, colored nodes denote terminal symbols, and edges represent grammar-driven expansion steps. This tree-structured perspective naturally supports tree-based search and learning algorithms.

\section{Reinforced Alpha Language Tree Search}


In the previous section, $\alpha$-Sem-$k$ induces a large yet well-structured tree of candidate alpha factors, where each leaf corresponds to a complete, evaluable expression.
Unlike conventional tree search problems, this space combines (i) explosive early branching, (ii) sharp contraction near a depth bound, and (iii) grammar-driven and formula-dependent actions, resulting in highly non-uniform search dynamics.
Moreover, the predictive performance of an alpha is revealed only at terminal nodes, which yields long-horizon dependencies and sparse rewards.

These challenges make unguided search ineffective and motivate a language-principled decision-making formulation.
Accordingly, we cast alpha discovery as a  Tree-Structured Linguistic Markov Decision Process  (TSL-MDP) and develop a reinforcement learning–guided, grammar-aware MCTS framework, supported by syntax-aware representation learning for efficient policy and value estimation.

\subsection{Decision-Making on Large Tree}
\label{subsec:mdp}
With \Cref{def:language_tree}, alpha discovery can be viewed as a sequential decision process over a large derivation tree. Equivalently, the task reduces  to: (i) selecting a high-quality root-to-leaf path that yields a strong alpha, or (ii) expanding an intermediate node (e.g., a partially specified or masked factor) into a more predictive expression. 

In this search tree, each complete alpha factor (leaf node) is evaluated by the average IC in \Cref{eq_icf}, computed from historical market data (\Cref{fig:tree search space}, \Cref{alg:incremental-combination}).
This  IC serves as the reward signal and can be propagated backward along the derivation path, assigning value estimates to intermediate nodes.
Consequently, partial expressions naturally correspond to states, grammar production rules to actions, and derivation steps to state transitions.

This perspective leads to a principled formulation of grammar-guided alpha discovery as a Markov Decision Process, which we term the \emph{Tree-Structured Linguistic Markov Decision Process (TSL-MDP)}.
\begin{definition}[TSL-MDP]
\label{def:TSL-MDP}
Alpha discovery under $\alpha$-Sem-$k$ is   a Tree-Structured Linguistic Markov Decision Process
$\text{TSL-MDP}=\langle S,A,P,R,\gamma\rangle$, where
$S$ is the set of partial or complete alpha expressions;
$A$ is the set of grammar production rules in  \Cref{df:alpha-CFG-semantic};
$P(s' \mid s,a)$ deterministically applies rule $a$ to expand the leftmost nonterminal in $s$, yielding a longer alpha expression $s'$;
and reward $R(s,a)$ is nonzero only when $s'$ is a complete alpha expression, equal to its IC evaluated on market data.
\end{definition}


\begin{figure*}[h] 
    \centering

    \includegraphics[width=0.85\textwidth]{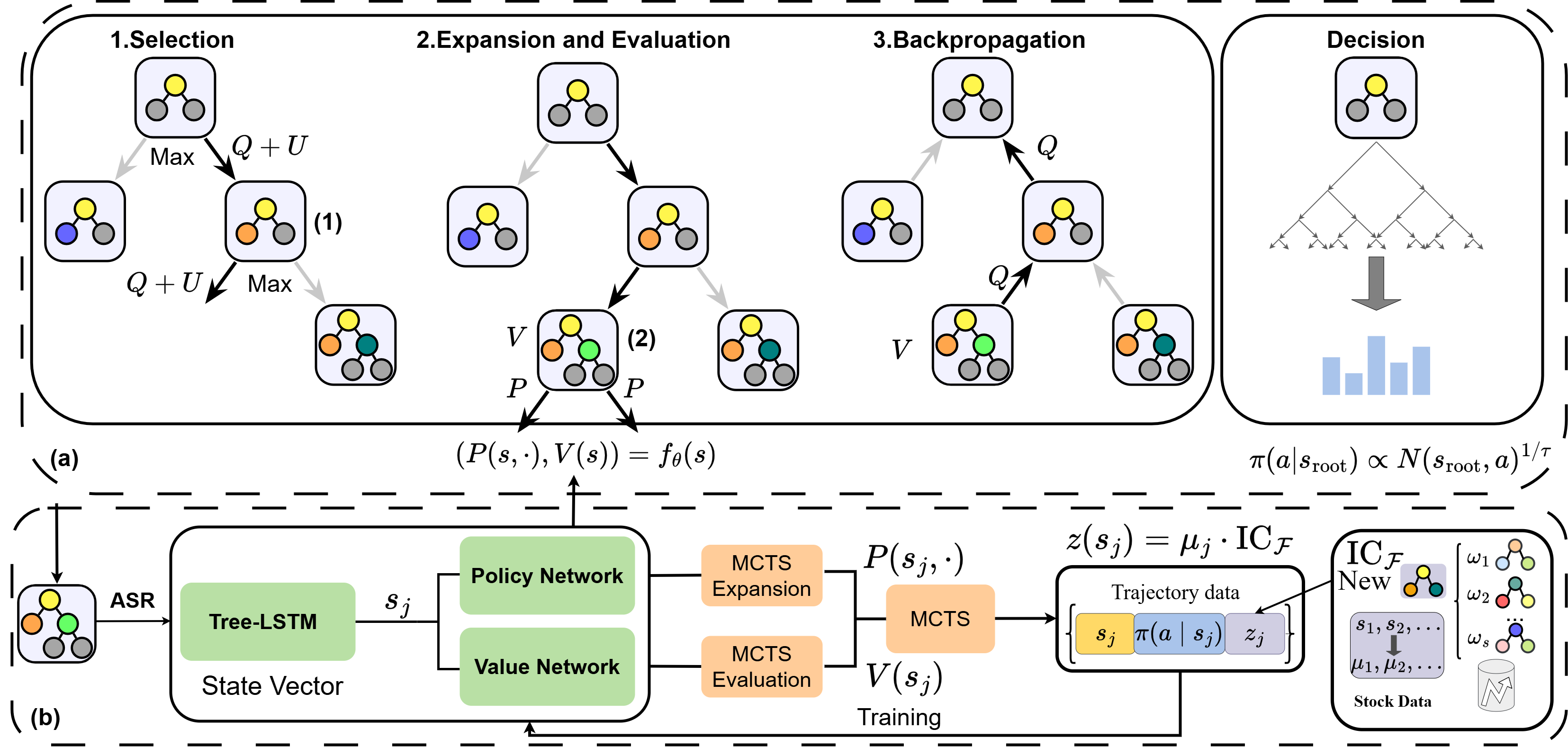}
    \caption{{Grammar-aware reinforcement learning and MCTS, based on alpha representation and value and policy networks.} } 
    \label{fig:mcts} 
\end{figure*}

\subsection{Reinforcement Learning Guided MCTS}
\label{sec:rl}




While the  tree structure of TSL-MDP makes it amenable to search, classical MCTS becomes ineffective at this scale due to long-horizon dependencies, highly irregular branching, and the absence of intermediate rewards.
We embed MCTS into a reinforcement learning framework that is explicitly tailored to grammar-based alpha generation.

Specifically, two neural networks are introduced: a policy network that predicts promising grammar production rules conditioned on a partial expression, and a value network that estimates the expected predictive quality of an incomplete alpha.
Both networks are driven by a Tree-LSTM encoder \citep{tai2015improved} that consumes the Abstract Syntax Representation (\Cref{def:expression tree}) of the current alpha expression, enabling structure-aware generalization across the vast TSL-MDP state space. Our framework is illustrated in \Cref{fig:frame}.

\paragraph{Overall Interaction Between RL and MCTS.}
Starting from the  start symbol of $\alpha$-Sem, alpha generation proceeds iteratively.
At each iteration $j$, we perform $I$ rounds of grammar-aware MCTS guided by the current policy and value networks.
The resulting search statistics induce a distribution over different production rules at the root, from which an action is sampled to expand to a node in the next layer of the search tree, which increases the current alpha expression.
The expanded node in this new layer then becomes the new root, and the process repeats until a complete alpha expression is generated.
Each completed alpha yields an IC reward, forming a trajectory of grammar decisions.
By collecting such trajectories, we iteratively update the policy and value networks via reinforcement learning, resulting in an effective \emph{search--learn--search} loop.
An overview of this interaction is illustrated in \Cref{fig:frame}, with the corresponding pseudocode provided in \Cref{alg:rl}.

\paragraph{MCTS Components.}
At a given root state $j$, the MCTS agent incrementally explores a subtree of the TSL-MDP through repeated simulations. Then it executes the following components (see \Cref{fig:mcts} (a) and Appendix \ref{Appendix-section-MCTS-full} for details).


\textit{Selection.} From the root, the MCTS agent repeatedly applies an $\alpha$-CFG production rule to the leftmost nonterminal symbol until reaching a frontier node which has not yet been expanded. 
The TSL-MDP exhibits highly irregular branching, with depth-dependent numbers of applicable production rules.
We therefore introduce an adaptive branching factor in the PUCT formulation, where $b$ denotes the number of valid actions at the current state and $b_{\text{ref}}$ is a normalization constant given by the maximum branching factor.
The ratio $\sqrt{\frac{b}{b_{\text{ref}}}}$ modulates the exploration term, emphasizing exploitation for small branching factors and promoting broader exploration for larger ones. Accordingly, we use the adapted PUCT-style selection rule~\citep{silver2017mastering}:
\begin{equation*}
\resizebox{\hsize}{!}{$
a^* = \arg\max_{a} \left(Q(s,a) + c_{\text{puct}}\sqrt{\tfrac{b}{b_\text{ref}}} \, P(s,a) \tfrac{\sqrt{\sum_b N(s,b)}}{1+N(s,a)} \right)
$}
\end{equation*}

\textit{Expansion and Evaluation}. 
Upon reaching a frontier node, all valid $\alpha$-CFG production rules are applied to generate its child states. The resulting node is evaluated using a Tree-LSTM–based value network $V(s)$, which estimates the expected terminal reward of the corresponding partial expression. Meanwhile, a policy network produces a distribution $P(s,a)$ over valid production rules, providing prior guidance for future selections.

\textit{Backpropagation}. The evaluation result $V(s)$ is backpropagated along the selection path, updating $Q(s,a)$ and visit counts $N(s,a)$. Iterating these steps allow MCTS agent progressively expands its explored subtree and refines the search statistics over the TSL-MDP (\Cref{alg:mcts-alpha-cfg}).

\begin{table*}[tb] 
  \small
  \centering
  \caption{Evaluation metrics comparison of different methods (5 random seeds).}
  \setlength{\tabcolsep}{0.5mm}
  {\scalebox{0.85}{
  \begin{tabular}{lcccccc}
    \toprule
    \multicolumn{7}{c}{\textbf{CSI300}} \\
    \midrule
    Method      & Rank IC               & IC                    & Rank ICIR            & ICIR                 & Sharpe               & Max Drawdown            \\
    \midrule
    XGBoost     & 0.0288 (0.0000)       & 0.0326 (0.0000)       & 0.2895 (0.0000)      & 0.2818 (0.0000)      & 0.2853 (0.0000)      & \textbf{-0.2777 (0.0000)} \\
    LightGBM    & 0.0539 (0.0029)       & 0.0296 (0.0014)       & 0.3963 (0.0247)      & 0.2649 (0.0395)      & 0.2680 (0.0666)      & -0.3271 (0.0177)         \\
    LSTM        & 0.0128 (0.0260)       & 0.0127 (0.0136)       & 0.0896 (0.2064)      & 0.1041 (0.1060)      & 0.1268 (0.0425)      & -0.3542 (0.0240)         \\
    TCN         & 0.0303 (0.0236)       & 0.0085 (0.0133)       & 0.2726 (0.1855)      & 0.0871 (0.1557)      & 0.0908 (0.0754)      & -0.2988 (0.0191)         \\
    ALSTM       & 0.0138 (0.0076)       & 0.0105 (0.0067)       & 0.1194 (0.0540)      & 0.0950 (0.0550)      & 0.1372 (0.1113)      & -0.3475 (0.0501)         \\
    Transformer    & 0.0423 (0.0133)       & 0.0248 (0.0132)       & 0.3759 (0.0697)      & 0.2457 (0.0971)      & 0.1699 (0.1105)      & -0.3365 (0.0377)         \\
    gplearn     & 0.0706 (0.0119)       & 0.0440 (0.0139)       & 0.4695 (0.1164)      & 0.3478 (0.1397)      & 0.2062 (0.2346)      & -0.3854 (0.0324)         \\
    AlphaQCM     & 0.0811 (0.0046)       & 0.0525 (0.0048)       & 0.5334 (0.0296)      & 0.3874 (0.0121)      & 0.4363 (0.0610)      & -0.3605 (0.0339)         \\
    
    RPN+PPO(AlphaGen)    & 0.0837 (0.0070)       & 0.0477 (0.0086)       & 0.5724 (0.0343)      & 0.3531 (0.0574)      & 0.4978 (0.1478)      & -0.3497 (0.0423)         \\

    \addlinespace
    \multicolumn{7}{l}{\textbf{Ablation Studies}} \\
    RPN+MCTS    & 0.0710 (0.0031)       & 0.0500 (0.0026)       & 0.5577 (0.0292)      & 0.4285 (0.0293)      & 0.5639 (0.1050)      & -0.3201 (0.0613)         \\
    $\alpha$-Syn+MCTS   & 0.0745 (0.0052)       & 0.0487 (0.0036)       & 0.5125 (0.0467)      & 0.3974 (0.0367)      & 0.4852 (0.1320)      & -0.3475 (0.0414)         \\
    $\alpha$-Sem+MCTS   & 0.0770 (0.0044)       & 0.0512 (0.0015)       & 0.5593 (0.0340)      & 0.4369 (0.0301)      & 0.5801 (0.1169)      & -0.3039 (0.0206)         \\

    \addlinespace
    $\alpha$-Sem-$k$+MCTS(AlphaCFG)        & \textbf{0.0865 (0.0060)} & \textbf{0.0577 (0.0029)} & \textbf{0.6036 (0.0537)} & \textbf{0.4505 (0.0249)} & \textbf{0.6459 (0.0612)} & -0.2963 (0.0289)         \\

    \midrule\midrule
    \multicolumn{7}{c}{\textbf{S\&P500}} \\
    \midrule
    Method      & Rank IC               & IC                    & Rank ICIR            & ICIR                 & Sharpe               & Max Drawdown            \\
    \midrule
    XGBoost     & 0.0140 (0.0000)       & 0.0104 (0.0000)       & 0.1535 (0.0000)      & 0.1456 (0.0000)      & 0.5883 (0.0000)      & -0.2543 (0.0000)       \\
    LightGBM    & 0.0078 (0.0021)       & 0.0220 (0.0032)       & 0.0860 (0.0269)      & 0.2072 (0.0229)      & 0.5852 (0.0547)      & -0.2047 (0.0128)         \\
    LSTM        & 0.0131 (0.0077)       & 0.0219 (0.0040)       & 0.1157 (0.0786)      & 0.1847 (0.0419)      & 0.5601 (0.0546)      & -0.2345 (0.0142)         \\
    TCN         & 0.0198 (0.0040)       & 0.0166 (0.0020)       & 0.1358 (0.0190)      & 0.1340 (0.0133)      & 0.4973 (0.0271)      & -0.2396 (0.0175)         \\
    ALSTM       & 0.0202 (0.0028)       & 0.0268 (0.0039)       & 0.1569 (0.0344)      & 0.1993 (0.0391)      & 0.4441 (0.0397)      & -0.2418 (0.0109)         \\
    Transformer    & 0.0106 (0.0049)       & 0.0185 (0.0036)       & 0.0828 (0.0433)      & 0.1806 (0.0361)      & 0.5979 (0.1163)      & -0.2512 (0.0070)         \\
    gplearn     & 0.0130 (0.0122)       & 0.0322 (0.0110)       & 0.0812 (0.0643)      & 0.1877 (0.0437)      & 0.8241 (0.1814)      & -0.2456 (0.0434)         \\
    AlphaQCM     & 0.0178 (0.0055)       & 0.0384 (0.0056)       & 0.1149 (0.0381)      & 0.2527 (0.0336)      & \textbf{1.0566 (0.0756)}      & -0.2105 (0.0273)         \\

    RPN+PPO(AlphaGen)    & 0.0149 (0.0055)       & 0.0342 (0.0050)       & 0.1045 (0.0364)      & 0.2420 (0.0296)      & 0.8271 (0.1421)      & -0.2559 (0.0242)         \\

    \addlinespace
    \multicolumn{7}{l}{\textbf{Ablation Studies}} \\
    RPN+MCTS      & 0.0309 (0.0054)       & 0.0385 (0.0031)       & 0.2447 (0.0234)       & 0.3308 (0.0344)      & 0.7992 (0.0854)      & -0.1957 (0.0140)         \\
    $\alpha$-Syn+MCTS   & 0.0111 (0.0017)       & 0.0272 (0.0047)       & 0.0913 (0.0087)      & 0.2335 (0.0356)      & 0.8046 (0.0322)      & -0.2286 (0.0186)         \\
    $\alpha$-Sem+MCTS  & 0.0265 (0.0011)       & 0.0413 (0.0030)       & 0.2075 (0.0108)     & 0.3360 (0.0162) & 0.8315 (0.0855) & -0.2243 (0.0225) \\

    \addlinespace
    $\alpha$-Sem-$k$+MCTS(AlphaCFG)        & \textbf{0.0354 (0.0026)} & \textbf{0.04573 (0.0034)} & \textbf{0.2958(0.0154)} & \textbf{0.4099 (0.0230)} & 0.8473 (0.0483) & \textbf{-0.1942 (0.0126)} \\
    \bottomrule
  \end{tabular}
  }}
  \label{tab:performance}
\end{table*}

\subsection{Syntax Representation Learning}
\label{sec:Representation Learning}

\label{sec:learnig}

\textbf{Network Design.} 
The main challenge in TSL-MDP is its vast state space, which requires evaluating both partial and complete alpha expressions as well as policies for expanding them. 
Since each state is naturally represented by an ASR (\Cref{def:expression tree}), we employ syntax-aware representation learning that directly encodes structure and semantics, which avoids costly rollout-based evaluations in classical MCTS.
Moreover, due to the symmetry of some operators (e.g., commutative operands), there are a large number of isomorphic factor expressions (defined in \Cref{def:tree_isomorphism}) in TSL-MDP. Syntax-aware representation learning is suitable for addressing these redundancies as it operates directly on ASRs rather than linear sequence. 

Accordingly, we use a Tree-LSTM encoder \citep{tai2015improved} with two heads: a policy head for predicting production-rule distributions and a value head for estimating terminal rewards (details are provided in Appendix~\ref{sec:Tree-lstm}).
As shown in \Cref{fig:mcts} (b), the Tree-LSTM recursively aggregates information, producing a fixed-dimensional state embedding for each ASR. This embedding is shared by both policy network for production-rule prediction and value network for state-value estimation in MCTS.

\textbf{Train and Sampling Procedure.} The policy and value networks are trained jointly using Tree-LSTM representations of TSL-MDP states. 
Initially, both networks are randomly initialized and used to guide MCTS expansion and evaluation. 
The resulting search statistics define an initial policy for alpha generation, which is then used to: (i) supervise the policy network via imitation of the MCTS-derived action distribution, and (ii) sample complete alpha expressions whose IC values (from market data) provide supervision for the value network. 
In subsequent iterations, the updated networks guide new MCTS constructions, and the process repeats until enough alphas have been sampled.

\textbf{Diversity-Aware Value Target.}
Since the ultimate objective is to construct a composite factor $\mathrm{IC}_{\mathcal{F}}$ (See \cref{alg:incremental-combination}), generating expressions that are structurally similar to existing factors can reduce pool diversity and degrade overall performance. To mitigate this, we incorporate a diversity-aware adjustment into the value target. 
Specifically, we define a normalized structural similarity measure $\mathrm{sim}(\cdot, \cdot)$, based on maximum common subtree matching \citep{sager2006tree} between the newly generated ASR $f_j$ corresponding to state $s_j$ and any existing $f_t \in \mathcal{F}$. This similarity penalizes states whose grammatical structures overlap with $\mathcal{F}$. The resulting value target is defined as
\begin{equation}
\label{eq:state_value}
z(s_j) = \bigl(1 - \max(0, \max_{f_t \in \mathcal{F}} \mathrm{sim}(f_t, f_j))\bigr) \cdot \mathrm{IC}_{\mathcal{F}}. 
\end{equation}
More details about tree similarity can be seen in \cref{sec:tree_similarity}.


\section{Experiments}

Detailed experimental settings, including datasets, comparison methods, evaluation metrics, and hyperparameters, are provided in the Appendix
(\Cref{sec:data}, \Cref{sec:Comparison Methods}, \Cref{sec:Evaluation Metrics}, and \Cref{sec:Parameters}).
Analysis on network architectures and mined factor examples with interpretability discussions are presented in
\Cref{sec:networks} and \Cref{sec:case study}, respectively.

\begin{figure}[htbp] 
  \centering
  \includegraphics[width=1\linewidth]{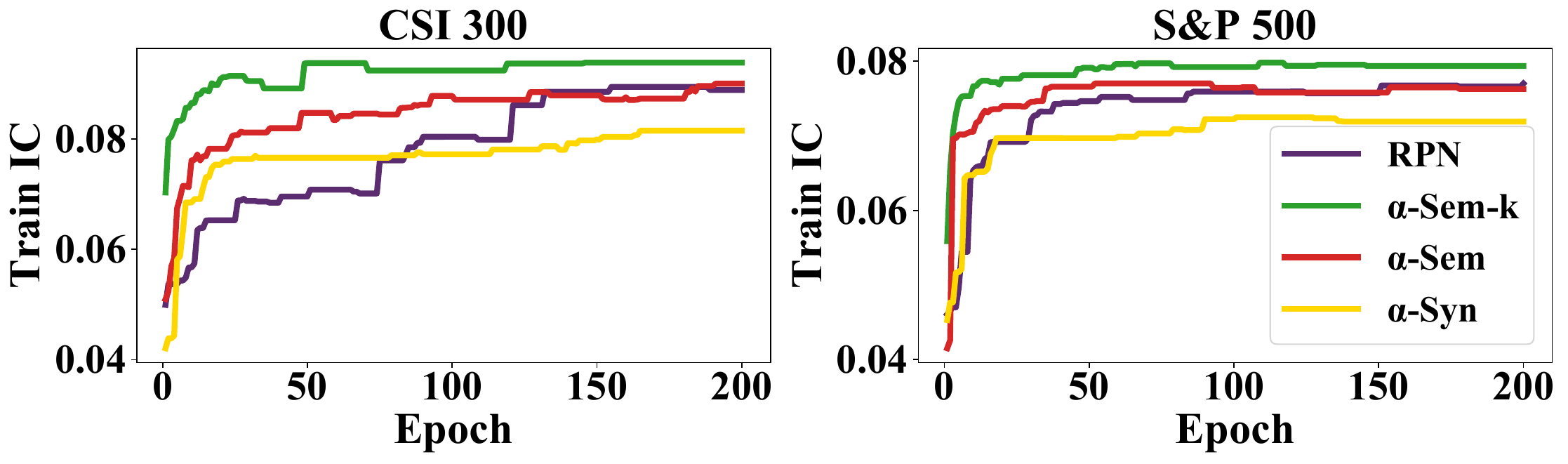}
  \caption{Comparison of training curves of generation methods.}
  \label{fig:train_ic}
\end{figure}

\textbf{Comparison of Generation Spaces.} We first compare different factor generation spaces (\Cref{fig:nested_ellipse}) to evaluate the impact of language constraints on factor discovery. Specifically, we compare three CFG levels with Reverse Polish Notation (RPN) \citep{krtolica2004reverse}, a computation and verification formalism with a non-recursive structure, on the CSI~300 and S\&P~500 training datasets.  
With a pool size of 10 and max length 5, \Cref{fig:train_ic} shows the training IC across epochs. 
Results confirm our analysis in \Cref{sec:language-level}, where more constrained, grammar-defined spaces yield faster convergence and higher-quality factors. 
Although RPN converges to a performance level close to $\alpha$-Sem, its convergence is noticeably slower.
This behavior reflects the limited semantic and length constraints of RPN, whose non-recursive structure restricts its effectiveness for structured factor generation compared to $\alpha$-Sem.

\begin{figure}[htbp]
    \centering
    \begin{subfigure}{0.4\textwidth}
        \centering
        \includegraphics[width=\linewidth]{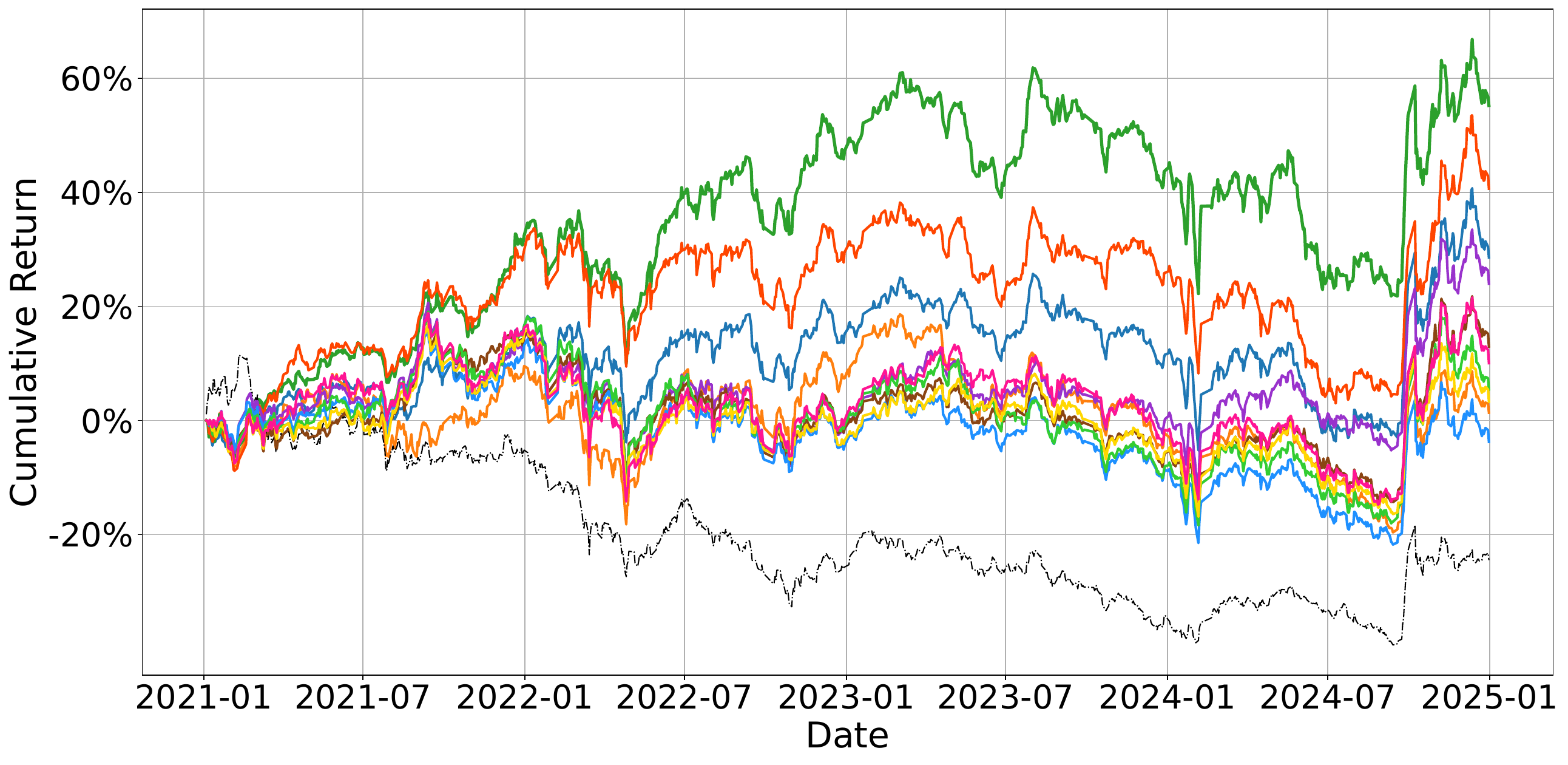}
        \caption{CSI 300}
        \label{fig:csi300}
    \end{subfigure}
    \hfill
    \begin{subfigure}{0.4\textwidth}
        \centering
        \includegraphics[width=\linewidth]{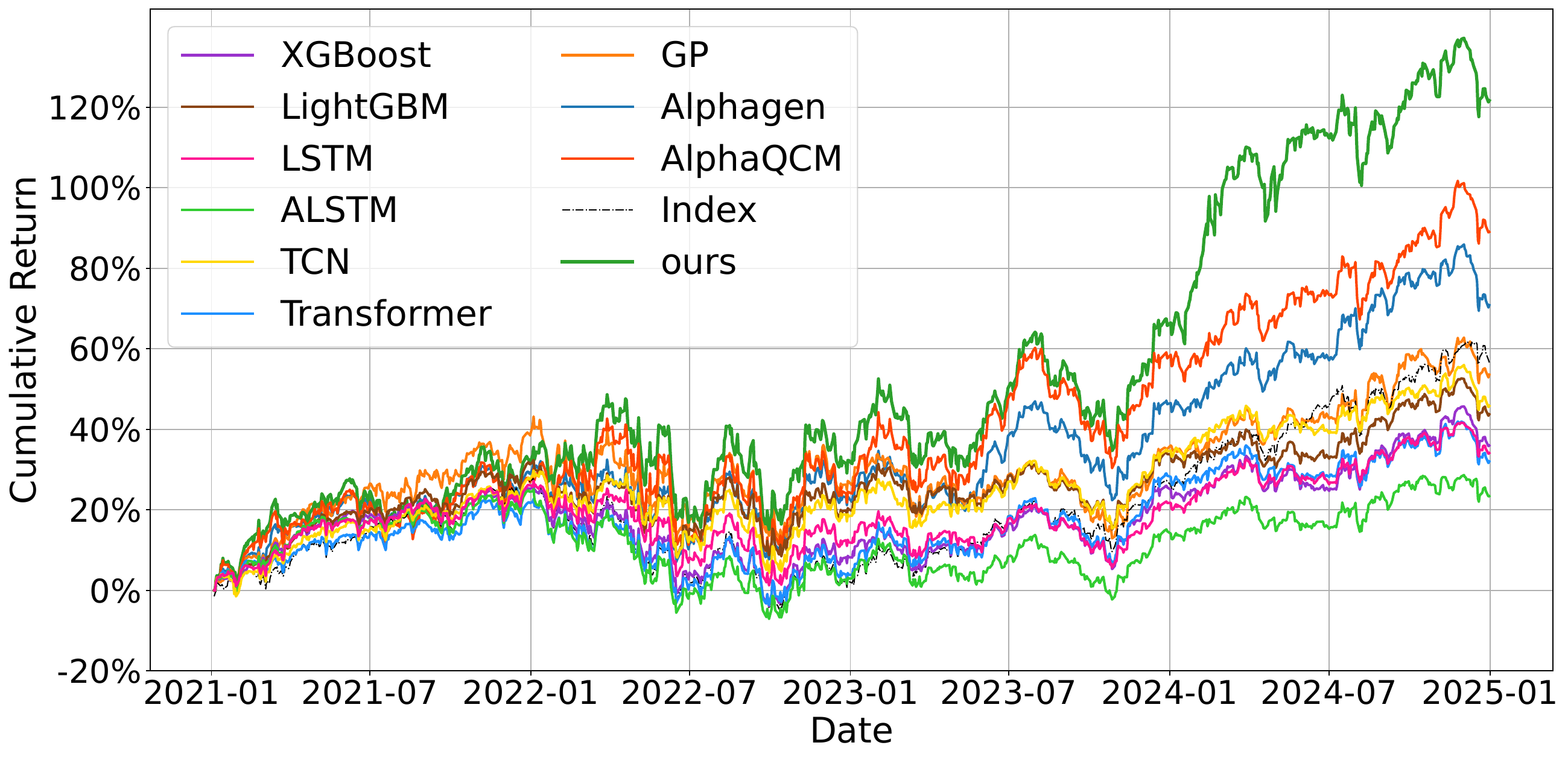}
        \caption{S\&P 500}
        \label{fig:sp500}
    \end{subfigure}
    \caption{Cumulative return comparison in simulated trading}
    \label{fig:comparison}
\end{figure}

\textbf{Comparison with Existing Alpha Mining Methods.} To create a fair comparison environment, we use the optimized hyperparameters from the validation dataset experiments (see details in Appendix \ref{sec:valid-Parameters}) for each method, including our MCTS-based methods ($\alpha$-Syn, $\alpha$-Sem, $\alpha$-Sem-$k$ and RPN) against existing factor mining methods or prediction models (formulaic: Alphagen, AlphaQCM, GPlearn; ML-based: XGBoost, LightGBM, LSTM, ALSTM, TCN, Transformer). The experiments were conducted separately on the CSI~300 index and the S\&P~500 constituents testing data, evaluating both correlation-based metrics and backtesting performance. The backtesting results are obtained using a single top-$k$/drop-$n$ strategy to conduct simulated trading based on real stock data (detailed in Appendix \ref{sec:Evaluation Metrics}). 
Quantitative results are summarized in \Cref{tab:performance}, and cumulative return curves are shown in \Cref{fig:comparison}.

Overall, our method achieves the best performance across all correlation metrics directly related to the optimization target IC.
Ablation studies further demonstrate the indispensable roles of syntactic constraints, semantic constraints, and length control.
While formulaic factor mining methods generally outperform machine-learning models that directly predict returns in correlation metrics, our approach also achieves strong backtesting performance.
Despite not directly optimizing for backtesting objectives, our method consistently attains superior Sharpe ratios and lower maximum drawdowns, and achieves the highest overall profitability among all compared methods.

\textbf{Improving Traditional Alpha Factors.}
Beyond directly mining composite factors, our $\alpha$-Sem-$k$+MCTS framework can be used to refine existing interpretable alpha factors.
We select a set of classic but recently ineffective factors from the GTJA 191 Factor Library and the Alpha101 Factor Library~\citep{kakushadze2016101}.
Factors from the GTJA 191 library are refined using the CSI~300 dataset, while Alpha101 factors are refined using the S\&P~500 dataset.
By partially masking operators and operands while preserving the left-side structure within half of the original expression length, we optimize these factors using a single-factor reward objective (illustrated by the blue path in \Cref{fig:tree search space}).
As shown in \Cref{tab:ic_comparison_revised}, our framework consistently improves the absolute IC values of many classic factors on the test datasets, demonstrating its effectiveness in strengthening existing alpha signals.

\begin{table}[htbp]
\centering
\caption{Refinement Results: Test Set IC Before and After Applying $\alpha$-Sem-$k$+MCTS framework.}
\label{tab:ic_comparison_revised}
\scriptsize
\begin{tabular}{@{} l r @{}}
\toprule
\multicolumn{2}{c}{\textbf{GTJA191}} \\
\midrule
Original: \textnormal{open/Ref(close,1)-1} & 0.00185 \\
Improved: \textnormal{open/0.1-Cov(volume,high,20)} & 0.04279 \\
\midrule
Original: \textnormal{Mean(close,6)-close} & 0.00482 \\
Improved: \textnormal{Mean(Cov(vwap,volume,20)/(-0.01),20)/0.05} & 0.04262 \\
\midrule
Original: \textnormal{close-Ref(close,5)} & 0.00495 \\
Improved: \textnormal{close-Greater(-0.1,Cov(volume,\textbar{}vwap\textbar{},30))} & 0.03872 \\
\bottomrule

\addlinespace[0.5em] 

\toprule
\multicolumn{2}{c}{\textbf{Alpha101}} \\
\midrule
Original: \textnormal{-Corr(open,volume,10)} & 0.00271 \\
Improved: \textnormal{Corr(open,Log(\textbar{}open\textbar{}),40)\textperiodcentered CSRank(high)} & 0.02934 \\
\midrule
Original: \textnormal{-Rank(CSRank(low),9)} & 0.01031 \\
Improved: \textnormal{Rank(CSRank(CSRank(Sign(vwap))),30)\textperiodcentered CSRank(high)} & 0.02944 \\
\midrule
Original: \textnormal{Pow(high\textperiodcentered\ low,0.5)-vwap)} & 0.00112 \\
Improved: \textnormal{Pow(CSRank(\textbar{}open\textbar{})\textperiodcentered open,CSRank(close))-vwap} & 0.03126 \\
\bottomrule
\end{tabular}

\end{table}

\section{Conclusion}
AlphaCFG formulates alpha factor discovery as a grammar-guided, syntax-tree–structured search problem that enforces interpretability while enabling efficient integration of reinforcement learning with neural Monte Carlo Tree Search.
Beyond trading, the framework naturally extends to other factor-based quantitative finance tasks.
More broadly, AlphaCFG exemplifies grammar-guided symbolic regression, where domain knowledge is encoded directly in the search space rather than learned implicitly from data.
A promising direction for future work is to integrate AlphaCFG with large-scale learned priors, such as foundation models over programs or syntax trees, to further accelerate search and improve generalization in structured reasoning problems.

\clearpage

\newpage


\bibliography{example_paper}
\bibliographystyle{icml2026}

\newpage
\appendix
\onecolumn

\section{Tables}

\begin{table}[!htbp]
    \centering
    \caption{Stock Feature Variables}
    \label{tab:stock_features}
    \scalebox{1}{
    \begin{tabular}{ll}
        \toprule
        \textbf{Feature} & \textbf{Description} \\
        \midrule
        $\text{open}$    & Opening price \\
        $\text{high}$    & Highest price \\
        $\text{low}$     & Lowest price \\
        $\text{close}$   & Closing price \\
        $\text{volume}$  & Trading volume \\
        $\text{vwap}$    & Volume Weighted Average Price (VWAP) \\
        \bottomrule
    \end{tabular}}
\end{table}

\begin{table}[!htbp]
    \centering
    \caption{Constant Parameters}
    \label{tab:constants}
    \begin{tabular}{ll}
        \toprule
        \textbf{Nonterminal} & \textbf{Values} \\
        \midrule
        $\text{Constant}$ & $-0.1,\,-0.05,\,-0.01,\,0.01,\,0.05,\,0.1$ \\
        $\text{Num}$      & $20,\,30,\,40$ \\
        \bottomrule
    \end{tabular}
\end{table}

\begin{table}[!htbp]
\centering
\caption{Formulaic Alpha Factor Operators in Our Framework (the BinaryOp in Formula~(\ref{fl-basic-production-rule}) does not distinguish whether it is symmetric)}
\scalebox{0.9}{
\begin{tabular}{>{\raggedright\arraybackslash}p{2.4cm} >{\raggedright\arraybackslash}p{2.2cm} p{7.8cm}} 
\toprule
\textbf{Operator} & \textbf{Type} & \textbf{Description} \\
\midrule
$\text{Abs}(x)$ & \textbf{Unary} & Absolute value, $\lvert x \rvert$. \\
$\text{Sign}(x)$ & \textbf{Unary} & Returns the sign of $x$: 1 for positive, -1 for negative, 0 for zero. \\
$\text{Log}(x)$ & \textbf{Unary} & Natural logarithm, $\log(x)$. \\
$\text{Add}(x, y)$ & \textbf{Binary} & Addition, $x + y$. \\
$\text{Mul}(x, y)$ & \textbf{Binary} & Multiplication, $x \cdot y$. \\
$\text{Greater}(x, y)$ & \textbf{Binary} & Returns the larger of two values: $\max(x, y)$. \\
$\text{Less}(x, y)$ & \textbf{Binary} & Returns the smaller of two values: $\min(x, y)$. \\
$\text{Div}(x, y)$ & \textbf{Binary-Asym} & Division, $x / y$. \\
$\text{Pow}(x, y)$ & \textbf{Binary-Asym} & Exponentiation, $x^y$. \\
$\text{Sub}(x, y)$ & \textbf{Binary-Asym} & Subtraction, $x - y$. \\
$\text{CSRank}(x)$ & \textbf{Rolling} & Cross-sectional ranking (normalizes the rank of $x$ across all stocks on the same day). \\
$\text{Rank}(x, t)$ & \textbf{Rolling} & Time-series ranking of $x$ over the past $t$ days. \\
$\text{WMA}(x, t)$ & \textbf{Rolling} & Weighted moving average with weights decaying over time. \\
$\text{EMA}(x, t)$ & \textbf{Rolling} & Exponential moving average with recursive smoothing. \\
$\text{Ref}(x, t)$ & \textbf{Rolling} & Value of $x$ from $t$ days ago. \\
$\text{Mean}(x, t)$ & \textbf{Rolling} & Mean of $x$ over the past $t$ days, $\frac{1}{t} \sum_{i=0}^{t-1} x_{-i}$. \\
$\text{Sum}(x, t)$ & \textbf{Rolling} & Sum of $x$ over the past $t$ days, $\sum_{i=0}^{t-1} x_{-i}$. \\
$\text{Std}(x, t)$ & \textbf{Rolling} & Standard deviation of $x$ over the past $t$ days. \\
$\text{Var}(x, t)$ & \textbf{Rolling} & Variance of $x$ over the past $t$ days. \\
$\text{Skew}(x, t)$ & \textbf{Rolling} & Skewness (measure of asymmetry) of $x$ over the past $t$ days. \\
$\text{Kurt}(x, t)$ & \textbf{Rolling} & Kurtosis (measure of tail thickness) of $x$ over the past $t$ days. \\
$\text{Max}(x, t)$ & \textbf{Rolling} & Maximum value of $x$ over the past $t$ days. \\
$\text{Min}(x, t)$ & \textbf{Rolling} & Minimum value of $x$ over the past $t$ days. \\
$\text{Med}(x, t)$ & \textbf{Rolling} & Median of $x$ over the past $t$ days. \\
$\text{Mad}(x, t)$ & \textbf{Rolling} & Mean absolute deviation, $\frac{1}{t} \sum_{i=0}^{t-1} \lvert x_{-i} - \bar{x} \rvert$. \\
$\text{Delta}(x, t)$ & \textbf{Rolling} & Difference, $x - \text{Ref}(x, t)$. \\
$\text{Cov}(x, y, t)$ & \textbf{PairedRolling} & Covariance between $x$ and $y$ over the past $t$ days. \\
$\text{Corr}(x, y, t)$ & \textbf{PairedRolling} & Pearson correlation coefficient between $x$ and $y$ over the past $t$ days. \\
\bottomrule
\end{tabular}
}
\vspace{-6pt}

\label{tab:operators}
\end{table}

\begin{table}[!h]
  \footnotesize
  \centering
   \caption{
   Length increments $\Delta k$ for each production rule.}
  \begin{tabular}{@{}>{\raggedright\arraybackslash}p{6cm} r@{}}
    \toprule
    \textbf{Production Rules} & $\Delta k$ \\
    \midrule
    \(\mathsf{Expr} \to \mathsf{Feature}\) & 0 \\
    \(\mathsf{Num} \to 20 \dots \) & 0 \\
    \(\mathsf{Constant} \to -0.01 \dots \) & 0 \\
    \(\mathsf{Expr} \to \mathsf{UnaryOp}(\mathsf{Expr})\) & 1 \\
    \(\mathsf{Expr} \to \mathsf{BinaryOp}(\mathsf{Expr}, \mathsf{Expr})\) & 2 \\
    \(\mathsf{Expr} \to \mathsf{BinaryOp}(\mathsf{Expr}, \mathsf{Constant})\) & 2 \\
    \(\mathsf{Expr} \to \mathsf{BinaryOp\_Asym}(\mathsf{Constant}, \mathsf{Expr})\) & 2 \\
    \(\mathsf{Expr} \to \mathsf{RollingOp}(\mathsf{Expr}, \mathsf{Num})\) & 2 \\
    \(\mathsf{Expr} \to \mathsf{PairedRollingOp}(\mathsf{Expr}, \mathsf{Expr}, \mathsf{Num})\) & 3 \\ 
    \bottomrule
  \end{tabular}
  \label{tab:delta-depth}
\end{table}

\newpage

\section{Algorithms}

\subsection{Linear combination alpha factor algorithm}\label{Appendix-section-Linear-Combination}
The linear combination factor model is defined as
\begin{equation}
c(X; F, w) = \sum_{j=1}^{n} w_j f_j(X) = y,
\label{eq:linear-combination}
\end{equation}
where $F = \{f_1, \dots, f_n\}$ denotes the set of factors, $w = \{w_1, \dots, w_n\}$ are the weights of factors in linear combination , $X$ represents the input stock feature data, and $y$ is the combined output. The optimization is conducted by minimizing the loss function
\begin{equation}
L(w) = \frac{1}{T} \sum_{t=1}^{T} \| y_t - r_t \|^2 
\label{eq:loss}
\end{equation}
where $r_t$ is the actual stock return, and $y_t$ is the alpha value of linear combination factor.  

\begin{algorithm}[h]
\caption{Incremental Combination Model Optimization}
\label{alg:incremental-combination}
\begin{algorithmic}
\STATE {\bfseries Input:} alpha set $F = \{f_1, \ldots, f_n\}$, weights $w = \{w_1, \ldots, w_n\}$, new alpha $f_{\text{new}}$
\STATE {\bfseries Output:} optimal alpha subset $F^{*}$, optimal weights $w^{*}$, combination IC $\mathrm{IC}_{\mathcal{F}}$

\STATE $F \gets F \cup \{ f_{\text{new}} \}$
\STATE $w \gets w \,\|\, \text{rand()}$

\FOR{$i = 1$ {\bfseries to} num\_gradient\_steps}
    \STATE Compute loss $L(w)$ according to Eq.~(\ref{eq:loss})
    \STATE $w \gets \text{GradientDescent}(L(w))$
\ENDFOR

\STATE $p \gets \arg\min_i |w_i|$
\STATE $F \gets F \setminus \{ f_p \}$; \quad $w \gets w \setminus \{ w_p \}$

\STATE Compute the combination IC: $\mathrm{IC}_{\mathcal{F}} \gets \text{IC}(F, w)$
\STATE {\bfseries Return} $F, w, \mathrm{IC}_{\mathcal{F}}$
\end{algorithmic}
\end{algorithm}

\subsection{Length control of semantic interpretable alpha factor generator}\label{appendix-section-length-control}
Following the intuition of grammar-constrained generation~\citep{jin2018unsupervised}, we introduce a \textit{$k$-bounded constraint} to explicitly limit expression length.
The mechanism maintains a counter $\mathit{k}$ for the partial length of the expression and enforces a maximum threshold $K$. Each production rule has a predefined increment $\Delta k$, representing its contribution to the expression length(see \Cref{tab:delta-depth} for details). A rule is applied only if
$
\mathit{k} + \Delta k \leq K,
$
thereby guaranteeing that each expansion step remains within the feasible bound.
By integrating this length-aware constraint into the derivation procedure, we obtain a bounded variant of $\alpha$-Sem, denoted as $\alpha$-Sem-k. 
The procedure is described in \Cref{alg:k-bounded-derivation}.

\begin{algorithm}[h]
  \caption{$\alpha$-Sem-$k$}
  \label{alg:k-bounded-derivation}
  \begin{algorithmic}
    \STATE {\bfseries Input:} Grammar $G = (\mathcal{N}, \mathcal{T}, \mathcal{P}, \mathcal{S})$, maximum length $K$, rule increments $\Delta k:  \Gamma \to \beta $
    \STATE {\bfseries Output:} Prefix expression tree $T$
    
    \STATE Initialize $T$ as a single-node tree with root $S$
    \STATE Set: $k \gets 0$
    
    \WHILE{$T$ contains a nonterminal node}
        \STATE Let $u$ be the first nonterminal node in a pre-order traversal of $T$
        \STATE Compute the set of applicable rules:
        \[
        \mathcal{A} \gets \{ l \in \mathcal{P} \mid l \text{ is applicable to } u \text{ and } k + \Delta k(l) \le K \}
        \]
        \STATE Choose rule $l: \Gamma \to \beta$ from $\mathcal{A}$
        \STATE Replace node $u$ with children corresponding to $\beta$
        \STATE Update: $k \gets k + \Delta k(l)$
    \ENDWHILE
    \STATE \textbf{Return} $T$
  \end{algorithmic}
\end{algorithm}

\newpage

\subsection{Algorithm of Four Stages of MCTS}\label{Appendix-section-MCTS-full}

\begin{algorithm}[h]
\caption{Grammar-aware MCTS with Branch-adapted PUCT}
\label{alg:mcts-alpha-cfg}
\begin{algorithmic}
\STATE {\bfseries Input:} root state $s_{\mathrm{root}}$, policy-value network $f_\theta$, iteration count $I$
\STATE {\bfseries Output:} improved policy $\pi(a \mid s_{\mathrm{root}})$

\FOR{$i = 1$ {\bfseries to} $I$}
    \STATE $s \gets s_{\mathrm{root}}$
    \STATE Initialize empty list of traversed edges $E \gets [\;]$

    \WHILE{$s$ is not fully expanded}
        \STATE $b \gets$ number of valid actions from $s$
        \STATE
        $
        a^* \gets
        \arg\max_a
        \Bigg[
            Q(s,a)
            + c_{\text{puct}}
            \cdot \sqrt{\tfrac{b}{b_{\mathrm{ref}}}}
            \cdot P(s,a)
            \cdot
            \tfrac{\sqrt{\sum_{b} N(s,b)}}{1 + N(s,a)}
        \Bigg]
        $
        \STATE Append $(s,a^*)$ to $E$
        \STATE $s \gets \text{apply}(s, a^*)$
    \ENDWHILE

    \STATE $s_L \gets s$
    \STATE $(P(s_L,\cdot), V(s_L)) \gets f_\theta(s_L)$ 
    \STATE Expand $s_L$ using $P(s_L,\cdot)$

    \FORALL{$(s,a) \in E$}
        \STATE $N(s,a) \gets N(s,a) + 1$
        \STATE
        $
        Q(s,a) \gets
        \frac{1}{N(s,a)}
        \sum_{s' \mid s,a \rightarrow s'} V(s')
        $
    \ENDFOR
\ENDFOR

\STATE
$
\pi(a \mid s_{\mathrm{root}})
=
\frac{N(s_{\mathrm{root}},a)^{1/T}}
{\sum_{b \in A(s_{\mathrm{root}})} N(s_{\mathrm{root}},b)^{1/T}}
$

\STATE {\bfseries Return} $\pi(a \mid s_{\mathrm{root}})$
\end{algorithmic}
\end{algorithm}

Assume that at a certain iteration $i$, our MCTS has already explored a portion of the TSL-MDP, denoted by an agent $M_i$. This agent corresponds to a subtree of the large  TSL-MDP, sharing the same root, and $M_i$ has  obtained policy  for this partial subtree. For example, at simulation $M_i$, the subtree agent $M_i$ shown on the left in \Cref{fig:mcts} has already been explored. This subtree starts as only a root when $i=0$, and is intended to expand toward the full TSL-MDP tree as $i$ increases, eventually reaching iteration $i=I$.

\textbf{Selection.} First, within $M_i$, starting from root of the subtree, the MCTS agent repeatedly selects an $\alpha$-CFG production rule at each incomplete alpha expression (each round-box node), and replaces its leftmost nonterminal symbol (the dark black arrows in \Cref{fig:mcts}), which goes to a new incomplete alpha expression (a child round-box node). This repeats until it reaches a ``frontier'' alpha expression that has a child not yet included in $M_i$ (e.g., node (1) in \Cref{fig:mcts}).

The TSL-MDP has two key features: (1) different nonterminal symbols have different numbers of production rules, and (2) the number of valid production rules decreases sharply near the bottom of the search tree due to the length control in  \ref{appendix-section-length-control}. To address this, we adopt a production rule selection function analogous to PUCT \citep{silver2017mastering}.
\begin{equation}
a^* = \arg\max_{a} \left( Q(s,a) + c_{\text{puct}} \cdot \sqrt{\tfrac{b}{b_\text{ref}}} \cdot P(s,a) 
\cdot \frac{\sqrt{\sum_b N(s,b)}}{1 + N(s,a)} \right),
\label{eq:puct}
\end{equation}
Here, $Q(s,a)$ is the value of selecting production rule $a$ for formula $s$, and  $P(s,a)$ is the probability of selecting $a$ under $s$.
$b$ is the number of branches at the current depth, and $b_\text{ref}$ is the branch balance constant (defined by the maximum number of branches)
Eq. (\ref{eq:puct}) balances irregular branching through the adaptive term $\sqrt{b/b_\text{ref}}$: smaller branching factors emphasize exploitation, while larger ones promote broader exploration.

\textbf{Expansion.} After finding such a frontier alpha expression node, the MCTS agent will execute a certain production rule on it, generating a new alpha expression which has not yet been covered by $M_i$ (e.g., round-box node (2) in \Cref{fig:mcts}), and also attaching all the corresponding possible production rules to this new alpha expression (e.g., the two arrows attached to node (2)).  
The probabilities for executing available production rules for expression $s$ follow  the distribution $P(s)$.

\textbf{Evaluation.} Since the newly expanded alpha expression is at the head of the current agent $M_t$ and remains incomplete, the existing policy cannot assess its quality. Thus, MCTS requires a method to evaluate it. Given the vastness of the TSL-MDP, traditional simulation-based evaluation is infeasible. Moreover, as shown in \Cref{def:expression tree}, the expressions at any state in TSL-MDP are small tree structures (i.e., the small trees inside each round-box in \Cref{fig:mcts}). Therefore, in the next section, we design a Tree-LSTM–based representation learning method to construct a value network for $V(s)$, as well as a policy network $P(s,a)$ over any expression.

\textbf{Backpropagation.} The result $V(s)$ of evaluation is backpropagated from the path of selection (the path directed by black arrow in the third tree of \Cref{fig:mcts}). Mean value of each edge in the path is updated by $V(s)$ and visit count $N(s,a)$ of each edge in the path increases by one.


The MCTS agent $M_i$ executes the above procedures at each iteration $i$ (\Cref{alg:mcts-alpha-cfg} shows the procedure of MCTS search.). Since one node is expanded at each step, the MCTS agent $M_i$ will eventually cover enough nodes and edges of the TSL-MDP. The resulting search assigns a \textit{basic value} to every node and obtain a basic policy for the TSL-MDP, which two can  be  used to further optimize the policy.

\section{Supplement to Problem Formulation}

\begin{figure*}[htbp]
  \centering

  \begin{subfigure}[t]{0.48\textwidth}
    \centering
    \includegraphics[width=\linewidth]{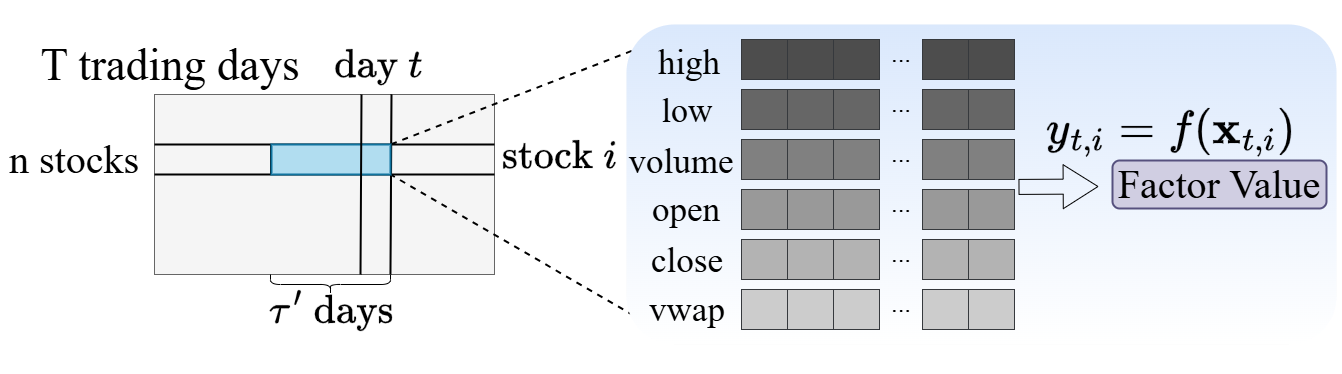}
    \caption{Illustration of an alpha factor.}
    \label{fig:alpha_factor}
  \end{subfigure}
  \hfill
  \begin{subfigure}[t]{0.48\textwidth}
    \centering
    \includegraphics[width=\linewidth]{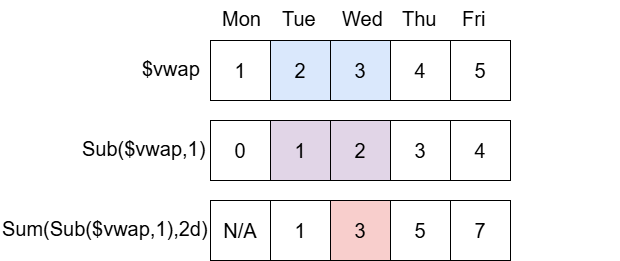}
    \caption{An example of a formulaic factor.}
    \label{fig:example}
  \end{subfigure}

  \caption{Alpha example.}
  \label{fig:alpha-example}
\end{figure*}

\Cref{fig:alpha_factor} illustrates the calculation process of an alpha factor. For a period of $T$ trading days, we compute the alpha factor for each stock using an alpha factor function 
\[
\mathbf{y}_t = (y_{t,1}, \dots, y_{t,n}) \in \mathbb{R}^n,
\] 
which takes as input the feature data of $n$ stocks over the current day $t$ and the previous $\tau'-1$ days. The resulting values represent the score of each stock for the current day, i.e., the alpha factor. These alpha values are subsequently used for stock selection and the formulation of trading strategies.

\Cref{fig:example} shows an example of formulaic factor: The factor $\mathrm{Sum}(\mathrm{Sub}(vwap, 1), 2d)$ computes the sum of the most recent two days of VWAP values after subtracting $1$ from each.
 To obtain the factor value on Wednesday, the operator first evaluates $\mathrm{Sub}(vwap,1)$ for Tuesday and Wednesday and then aggregates them:
     $(2-1) + (3-1) = 3$.
     This output serves as the alpha signal, the predicted return for Wednesday which is subsequently used in downstream stock-selection or portfolio-construction procedures.

\newpage
\section{Reinforcement Learning Framework}

We present pseudo-code of MCTS combined with reinforcement learning method (\Cref{alg:rl}). This is a reinforcement learning-based factor mining method designed to automatically discover a combination of factors from stock market data that can effectively predict stock returns. Specifically, the algorithm initializes a set of factors, their corresponding weights, and a policy-value network. In the process of obtaining data through reinforcement learning, it employs a MCTS policy to generate actions for each state, thereby constructing a multi-step factor generation path. The final state of the path is parsed into a computable alpha expression, evaluated using the $IC_{\mathcal{F}}$ as the reward signal. The reward is given along with the optimization of the factor combination $\mathcal{F}$. The actual value for each step along the path, denoted as $z_t$ is computed based on $IC_{\mathcal{F}}$ and the similarity between the newly generated factor and existing ones, following the formulation in \Cref{eq:state_value} in \Cref{sec:learnig}. 

After generating multi-step factor paths in each iteration, the policy and value networks are trained using the collected path data $(s_j, \pi(a|s_j), z_j)$ stored in a replay buffer, where $s_j$ is the state vector encoded by TreeLSTM, $\pi(a|s_j)$ is the policy from MCTS, and $z_t$ is shown above. After training, the networks are redeployed to guide a new round of search. Through iterative training and exploration, the IC of the learned factor combination is progressively improved. The algorithm outputs the final optimized factor combination set along with its corresponding weights when the IC shows no more significant improvement.

\begin{figure*}[!htbp] 
    \centering
    \includegraphics[width=0.8\linewidth]{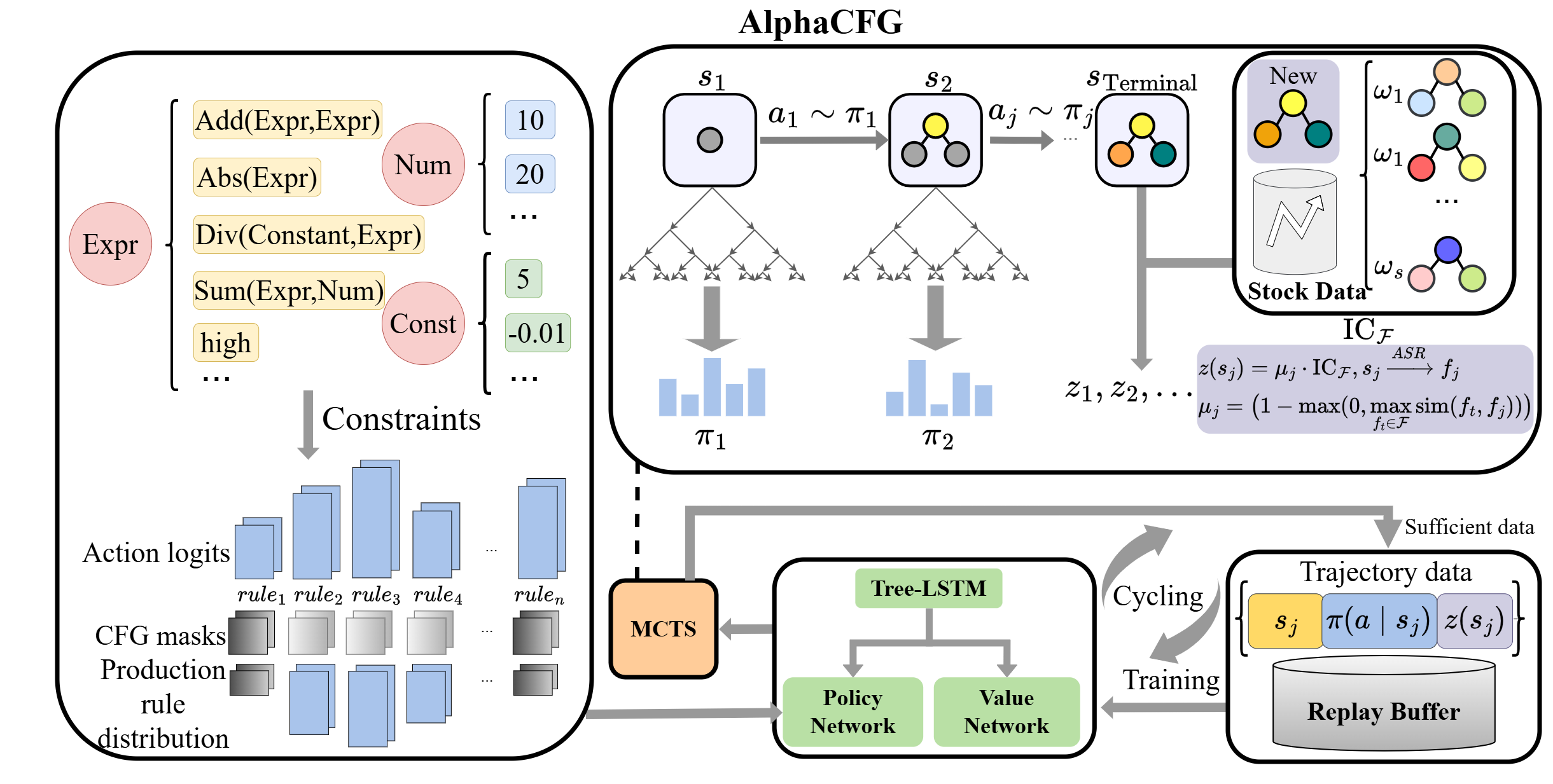}
    \caption{The overall framework of AlphaCFG.}
    \label{fig:frame} 
\end{figure*}

The overall workflow of this algorithm is illustrated in \Cref{fig:frame} in the following page, while a specific illustration of its MCTS component \Cref{alg:mcts-alpha-cfg} is in \Cref{fig:mcts}, and the illustration of its neural network part is in \Cref{fig:mcts} (b).

\begin{algorithm}[h]
\caption{Alpha Mining via Reinforcement Learning}
\label{alg:rl}
\begin{algorithmic}
\STATE {\bfseries Input:} stock trend dataset $Y = \{y_t\}$
\STATE {\bfseries Output:} optimal alpha subset $F^*$, optimal weights $w^*$

\STATE Initialize alpha set $F$ and weights $w$
\STATE Initialize policy-value network $f_\theta$ and replay buffer $D$

\FOR{each epoch}
    \FOR{each factor path search}
        \STATE Initialize empty trajectory $E \gets [\;]$

        \FOR{$j = 0$ {\bfseries to} $J$}
            \STATE Append state $s_j$ to $E$
            \STATE $s_{\mathrm{root}} \gets s_j$
            \STATE
            $
            \pi(a \mid s_j) \gets \pi(a \mid s_{\mathrm{root}})
            $
            \STATE Sample action $a_j \sim \pi(a \mid s_j)$
            \STATE $s_{j+1} \gets \text{apply}(s_j, a_j)$
        \ENDFOR

        \STATE $f_j \gets \text{parse}(s_{J})$
        \STATE Obtain reward $\mathrm{IC}_{\mathcal{F}}$ using \cref{alg:incremental-combination}

        \STATE {\bfseries Reward Assignment}
        \FOR{$j = 0$ {\bfseries to} $J$}
            \STATE
            $
            z(s_j) \gets
            \Bigl(
            1 - \max\bigl(0, \max_{f_t \in F} \mathrm{sim}(f_t, f_j)\bigr)
            \Bigr)
            \cdot
            \mathrm{IC}_{\mathcal{F}}
            $
            \STATE $D \gets D \cup \{(s_j, \pi(a \mid s_j), z(s_j))\}$
        \ENDFOR
    \ENDFOR

    \STATE {\bfseries Network Update}
    \FOR{each gradient step}
        \STATE Sample minibatch $B \subset D$
        \STATE
        $
        L_\theta =
        \left(z(s_t) - V_\theta(s_t)\right)^2
        -
        \sum_a \pi(a \mid s_t)\log P_\theta(a \mid s_t)
        +
        c \|\theta\|^2
        $
        \STATE $\theta \gets \theta - \eta \nabla_\theta L_\theta$
    \ENDFOR
\ENDFOR

\STATE {\bfseries Return} $F^*, w^*$
\end{algorithmic}
\end{algorithm}

\section{Search Space Complexity}
\label{sec:analysis}

To compare the sizes of expression search spaces under different generation methods, we study three methods from a combinatorial perspective: 
(i) a purely exponential baseline (arbitrary combination of all symbols corresponding to $\Sigma^*$); 
(ii) $\alpha$-Syn (corresponding to $\mathcal{L}_{\mathrm{syn}}$); 
(iii) $\alpha$-Sem (corresponding to $\mathcal{L}_{\mathrm{syn}}$).  
All three methods share the same parameter sets (operator types, number of features, constants, etc.), but progressively impose stricter constraints, resulting in smaller search spaces.

We set the following notation: the size of the unary operator set is $|U|$, the size of the binary operator set is $|B|$, the size of the asymmetric binary operator set is $|B_{\text{asym}}|$, the size of the rolling operator set is $|R|$, the size of the paired rolling operator set is $|R_{\text{pair}}|$, the number of features is $|\mathcal{F}|$, the number of constant parameters is $|\mathcal{C}|$, and the number of rolling-window parameters is $|\mathcal{N}|$.


\subsection{Unstructured Space $\Sigma^*$}

The method of arbitrary symbol combination (referred to ) takes one symbol equally at each step from all available symbols.  
Let the total number of symbols be:
\[
r = |\mathcal F| + |\mathcal C| + |\mathcal N|
      + |U| + |B| + |B_{\text{asym}}| + |R| + |R_{\text{pair}}|.
\]

Then the number of sequences of length $n$ is
$
r_n = r^n,
$
and the cumulative size is
$
\sum_{i \le n} r_i = \Theta(r^n).
$

\subsection{Syntactically Legal Space $\mathcal{L}_{\mathrm{syn}}$}

We introduce syntax constraints to ensure that generated expressions are all syntactically valid.  
We consider the grammar $\alpha$-Syn:

\[
\mathsf{Expr} \rightarrow
\mathsf{UnaryOp}(E)\mid
\mathsf{BinaryOp}(E,E)\mid
\mathsf{RollingOp}(E,E)\mid
\mathsf{PairedRollingOp}(E,E,E)\mid
\mathsf{TermSyb}.
\]

Let $h_n$ be the number of valid expressions of length $n$.  
The terminal set size is:
$
T = |\mathcal F| + |\mathcal C| + |\mathcal N|.
$

Define operator cardinalities:
$
U = |U|,
Q = |B| + |B_{\text{asym}}|,
R = |R|,
P = |R_{\text{pair}}|
$, respectively(The meanings of the notations are as shown in D).

The recurrence formula is:
$
h_1 = T,
$
and for $n \ge 2$:
\[
h_n
= U h_{n-1}
+ (Q+R)\sum_{i=1}^{n-2} h_i \, h_{n-1-i}
+ P \!\!\!\sum_{\substack{i+j+k=n-1 \\ i,j,k\ge 1}} h_i h_j h_k.
\]
The subsequent derivation of an explicit form from this recurrence becomes rather cumbersome. Since the technical steps mirror the usual treatment of general cubic functional equations, we omit the full derivation here. 

\subsection{Semantically Legal Space $\mathcal{L}_{\mathrm{sem}}$}

$\alpha$-Sem introduces more constraints on constants, argument types, and rolling windows:

\[
\begin{aligned}
\mathsf{Expr} \to &
\ \mathsf{Feature}
\mid \mathsf{UnaryOp}(\mathsf{Expr})
\\ &\mid \mathsf{BinaryOp}(\mathsf{Expr},\mathsf{Expr})
\mid \mathsf{BinaryOp}(\mathsf{Expr},\mathsf{Constant})
\\ &\mid \mathsf{BinaryOp\_Asym}(\mathsf{Constant},\mathsf{Expr})
\mid \mathsf{RollingOp}(\mathsf{Expr},\mathsf{Num})
\\ &\mid \mathsf{PairedRollingOp}(\mathsf{Expr},\mathsf{Expr},\mathsf{Num}),
\\[4pt]
\mathsf{Num} \to &\ 20 \mid \cdots,
\qquad
\mathsf{Constant} \to -0.01 \mid \cdots
\end{aligned}
\]

Let $f_n$ denotes the number of valid expressions of length $n$.

The recurrence formula becomes
\[
\begin{aligned}
f_{n} =\;&
|U|\, f_{n-1} &&
\text{(unary)} \\[6pt]
&+ |B| \sum_{i=1}^{n-2} f_i f_{n-1-i} &&
\text{(binary)} \\[6pt]
&+ |B|\, |\mathcal C| \, f_{n-2} &&
\text{(binary + right constant)} \\[6pt]
&+ |B_{\text{asym}}|\, |\mathcal C| \, f_{n-2} &&
\text{(asymmetric binary + left constant)} \\[6pt]
&+ |R|\, |\mathcal N| \, f_{n-2} &&
\text{(rolling)} \\[6pt]
&+ |R_{\text{pair}}|\, |\mathcal N| 
   \sum_{i=1}^{n-3} f_i f_{n-2-i} &&
\text{(paired rolling)}.
\end{aligned}
\]

The recurrence formula is similar, and compared with $\alpha$-Syn, recurrence of $\alpha$-Sem includes more convolution terms and more realistic constraints, providing a more accurate operator usage. In the following, we present the overall analysis.

Because the expression length is unbounded, the search spaces of all three generation methods are infinite.
 Therefore, the comparison does not concern the total size of each space, but rather the size of the finite subspace consisting of expressions whose length is at most $ n $.

For each grammar, the production rules yield a recurrence for the number of expressions of exact length $n$  ) ( $r_n, h_n, f_n $), and accumulating these values from $ 1$  to  $n$  gives the size of the corresponding truncated subspace. By computing these cumulative counts and plotting their growth as functions of $n$, we can directly compare how quickly the reachable portions of the three search spaces expand.

\subsection{Empirical Verification}

\begin{figure}[h]
    \centering
    \includegraphics[width=0.70\textwidth]{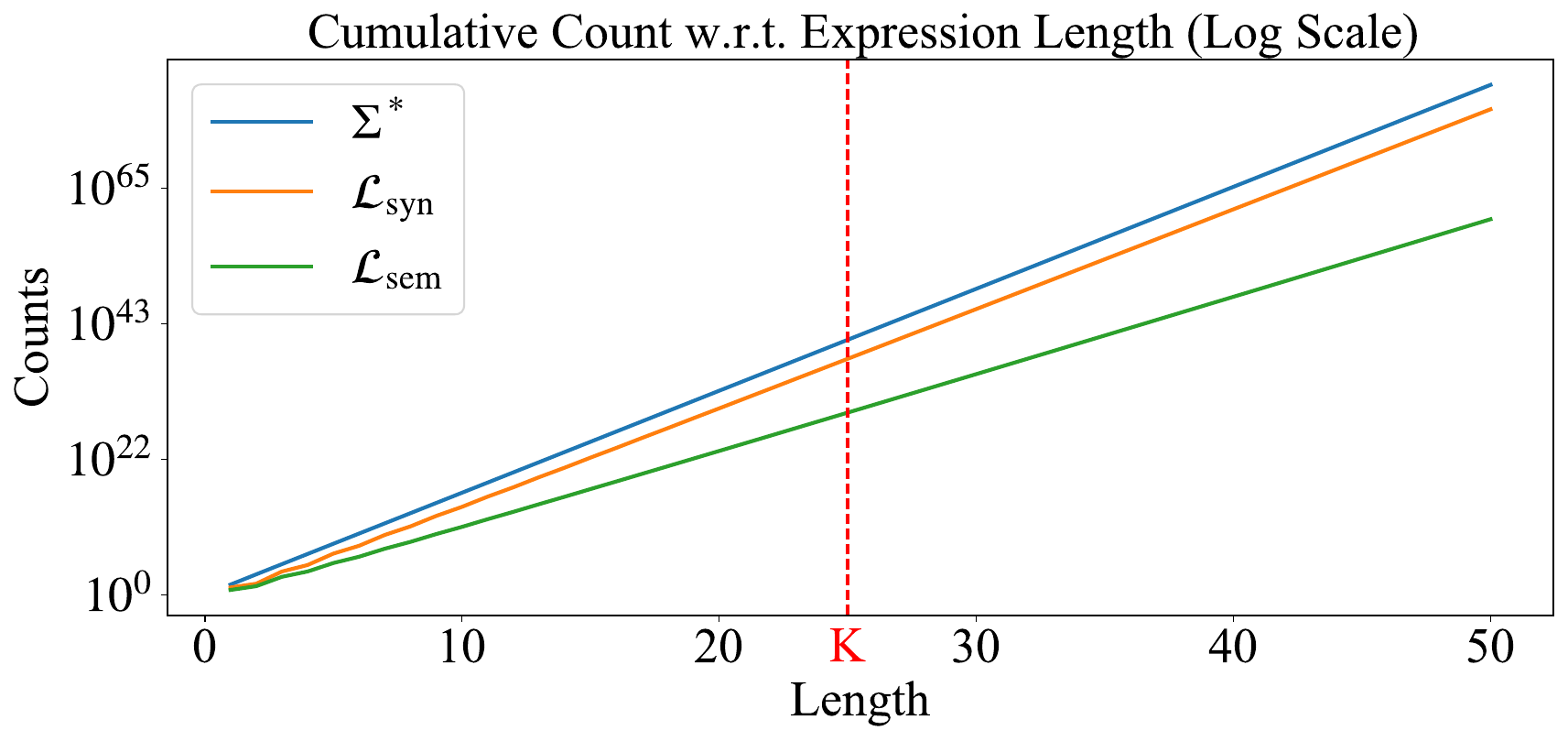}
    \caption{Comparison of cumulative search space sizes of different grammar levels.}
    \label{fig:space}
\end{figure}

Based on the recurrence formulas, We compute the cumulative counts of $\{r_n\}$, $\{h_n\}$, and $\{f_n\}$ for $n=1 \sim$  $N$, and plot their growth curves to visualize differences between the three methods (shown in \Cref{fig:space}). Since all three methods yield inherently infinite search spaces, we further design $\alpha$-Sem-k based on \Cref{alg:k-bounded-derivation}, which can be seen as the red dotted line in \Cref{fig:space}. The results are consistent with the analysis in \Cref{fig:nested_ellipse}, which further strengthens the superiority of our approach in theory. 

\Cref{fig:space} explains the core of the superiority of our method: By introducing constraints of syntax and semantics, We get an infinite set containing only valid factors. In actual factor search tasks, we cannot exhaust this space that exploring a finite subset is realistic. Therefore, We utilize the recursive feature of CFG and further designed $\alpha$-Sem-k capable of generating factors of only a finite length. Ultimately, we reduced the complexity of the search space from an exponential level to a constant level, making this task solvable.

\section{Details of Tree-LSTM}
\label{sec:Tree-lstm}


%
Starting from ASR leaf nodes, the Tree-LSTM recursively aggregates child hidden and cell states through gating (input, forget, output), combining them with the node’s input embedding. This bottom-up process continues until the root, yielding a fixed-dimensional state vector that encodes both the syntax and operator-specific dependencies of the entire expression. Thus, the Tree-LSTM transforms variable-sized trees into single vectors while preserving structural and semantic information.

In our $\alpha$-CFG, operators are different: (\textit{i}) {symmetric operators}, where order is irrelevant, and (\textit{ii}){asymmetrical (order-sensitive) operators}, where order must be preserved. 
Tree-LSTM naturally supports both cases through two variants: the N-ary Tree-LSTM, which uses position-sensitive parameters to encode child order, and the Child-Sum Tree-LSTM, which aggregates child states by their mean to provide order-invariant representations. 
Based on these, we tailor aggregation strategies: for symmetric binary operators (\(\mathsf{Expr} \to \mathsf{BinaryOp}(\mathsf{Expr}, \mathsf{Expr})\)) we adopt Child-Sum to avoid redundant encodings; for paired rolling operators (\(\mathsf{Expr} \to \mathsf{PairedRollingOp}(\mathsf{Expr}, \mathsf{Expr}, \mathsf{Num})\)) we first apply unordered aggregation to operands and then use N-ary encoding to incorporate the time-window parameter; and for all other operators we employ standard N-ary encoding. Such operation can address the problem of isomorphic redundancy of alpha factors defined in \Cref{def:tree_isomorphism}.The resulting tree embeddings are treated as input to be given into the policy and value heads to predict next-rule probabilities and estimated state value.

\subsection{N-ary Tree-LSTM (Position-Sensitive)}
  
Let node $j$ have $N$ children with hidden states $\mathbf{h}_1, \dots, \mathbf{h}_N$, input $\mathbf{x}_j$, output hidden state $\mathbf{h}_j$ and cell state $\mathbf{c}_j$:
{\small{\begin{equation}
\begin{aligned}
\mathbf{i}_j &= \sigma\left(W^{(i)} \mathbf{x}_j + \sum_{k=1}^{N} U_k^{(i)} \mathbf{h}_k + \mathbf{b}^{(i)} \right) \\
\mathbf{f}_{jk} &= \sigma\left(W^{(f)} \mathbf{x}_j + U_k^{(f)} \mathbf{h}_k + \mathbf{b}^{(f)} \right), \quad k=1,\dots,N\\
\mathbf{o}_j &= \sigma\left(W^{(o)} \mathbf{x}_j + \sum_{k=1}^{N} U_k^{(o)} \mathbf{h}_k + \mathbf{b}^{(o)} \right) \\
\mathbf{u}_j &= \tanh\left(W^{(u)} \mathbf{x}_j + \sum_{k=1}^{N} U_k^{(u)} \mathbf{h}_k + \mathbf{b}^{(u)} \right) \\
\mathbf{c}_j &= \mathbf{i}_j \odot \mathbf{u}_j + \sum_{k=1}^{N} \mathbf{f}_{jk} \odot \mathbf{c}_k \\
\mathbf{h}_j &= \mathbf{o}_j \odot \tanh(\mathbf{c}_j)
\end{aligned}\nonumber
\end{equation}}}

\subsection{Child-Sum Tree-LSTM}

Let node $j$ have a set of children $C(j)$ with hidden states $\mathbf{h}_k$, $k \in C(j)$:

\[
\begin{aligned}
\tilde{\mathbf{h}}_j &= \frac{1}{|C(j)|} \sum_{k \in C(j)} \mathbf{h}_k \\
\mathbf{i}_j &= \sigma\left(W^{(i)} \mathbf{x}_j + U^{(i)} \tilde{\mathbf{h}}_j + \mathbf{b}^{(i)}\right) \\
\mathbf{f}_{jk} &= \sigma\left(W^{(f)} \mathbf{x}_j + U^{(f)} \mathbf{h}_k + \mathbf{b}^{(f)} \right), \quad k \in C(j) \\
\mathbf{o}_j &= \sigma\left(W^{(o)} \mathbf{x}_j + U^{(o)} \tilde{\mathbf{h}}_j + \mathbf{b}^{(o)}\right) \\
\mathbf{u}_j &= \tanh\left(W^{(u)} \mathbf{x}_j + U^{(u)} \tilde{\mathbf{h}}_j + \mathbf{b}^{(u)}\right) \\
\mathbf{c}_j &= \mathbf{i}_j \odot \mathbf{u}_j + \sum_{k \in C(j)} \mathbf{f}_{jk} \odot \mathbf{c}_k \\
\mathbf{h}_j &= \mathbf{o}_j \odot \tanh(\mathbf{c}_j)
\end{aligned}
\]

\section{Calculation of Tree Similarity}
\label{sec:tree_similarity}

\begin{definition}[Isomorphism of ASR(Tree)]
\label{def:tree_isomorphism}
ASR \(T_1\) and \(T_2\) are isomorphic only if:
\begin{enumerate}
    \item The label of root nodes must be the same;
    \item Recursively check each child node, the labels of the child nodes are equivalent: for asymmetrical operations, the order of the subtrees must be preserved; for symmetrical operations (Binary type operators in  \Cref{tab:operators}) or partially symmetrical operations (Corr, Cov, where the order of the first two operands' child nodes doesn't matter), the order of the subtrees doesn't matter as long as the operands match;
    \item Recursively check that all child nodes and their structures are isomorphic.
\end{enumerate}
\end{definition}

Given two alpha factor expresions(partial or completed), they correspond to two ASRs \(T_1\) and \(T_2\) which are also two trees. Let \(\text{Sub}(T)\) denote the set of all subtrees of \(T\), where each subtree is induced by a child of node in \(T\) along with all its descendant nodes (including the child node itself). Let \(N(T)\) denote the total number of subtrees in \(T\), recursively defined as:
\[
N(T) = 1 + \sum_{c \in \text{Children}(T)} N(c).
\]

The normalized similarity between the two ASR is defined as:
\[
\mathrm{sim}(T_1, T_2) = \frac{\max_{\substack{t_1 \in \text{Sub}(T_1) \\ t_2 \in \text{Sub}(T_2)}} \mathrm{css}(t_1, t_2)}{\max\left(N(T_1),\; N(T_2)\right)},
\]
where the numerator represents the size of the largest isomorphic subtree shared by \(T_1\) and \(T_2\), i.e., the number of matching nodes in the largest common subtree. Tree isomorphism is defined formally in ~\Cref{def:tree_isomorphism}. If no such isomorphic subtree exists, then \(\mathrm{css}(t_1, t_2) = 0\).

The denominator \(\max(N(T_1), N(T_2))\) corresponds to the number of nodes in the larger of the two trees, serving as an upper bound for the size of any common subtree. Intuitively, it reflects the maximum number of matching nodes that could be achieved if one tree were a subtree of the other, or if the two trees were structurally identical. As such, the denominator defines the \emph{maximum potential scale} of a common subtree, and serves to normalize the matching node count in the numerator. This ensures that the resulting similarity score lies within the standardized range \([0, 1]\), thereby facilitating both quantitative analysis and intuitive comparison of structural similarity between expression trees.



\section{AlphaCFG Framework Parameter Setting for Experiment}
\label{sec:Parameters}

\subsection{MCTS Parameters}
\begin{itemize}
    \item Exploration Parameter : The exploration-exploitation trade-off parameter in the UCT formula is set to \( c = 1 \).

    \item MCTS Simulations : 64 simulations are performed per state.

    \item MCTS Parallelism: 8 parallel simulations are used to speed up the exploration.

    \item Eval Batch Size: 2 evaluations using network are carried out simultaneously each time.

    \item Branch balance coefficient: 40
    
\end{itemize}

\subsection{Network Architecture}

\textbf{Feature Extractor (Tree-LSTM)}:
\begin{itemize}
    \item Embedding Dimension: 128.
    \item Hidden Size: 128.
    \item Dropout Rate: 0.1.
\end{itemize}

\textbf{Policy Network}: 
\begin{itemize}
    \item Input: Features extracted by the feature extractor (Tree-LSTM).
    \item Hidden Layers: 
    \begin{itemize}
        \item Layer 1: Fully connected layer with 128 input features and 64 output features.
        \item Layer 2: Fully connected layer with 64 input features and 128 output features (embedding dimension).
    \end{itemize}
    \item Activation Function: Softmax
\end{itemize}

\textbf{Value Network}: 
\begin{itemize}
    \item Input: Features extracted by the feature extractor (Tree-LSTM).
    \item Hidden Layers:
    \begin{itemize}
        \item Layer 1: Fully connected layer with the embedding dimension (128) as input and 64 output features.
        \item Layer 2: Fully connected layer with 64 input features and 64 output features.
    \end{itemize}
    \item Activation Functions: ReLU activation functions applied to the hidden layers.
    \item Output: A fully connected layer with a single output value without activation function. 
\end{itemize}

\subsection{Optimizer and Training Parameters}
\begin{itemize}
    \item Optimizer: Adam optimizer with default settings
    \item Learning Rate: A learning rate of \( 10^{-4} \).
    \item Batch Size: 64.
    \item Number of factor trajectories in an iteration: 100(2*50).
    \item Training Iterations: 100 iterations.
    \item Batch Size for Training: 64.
    \item Replay Buffer Size: 20,000.
    \item Early Stopping Criteria: Early stopping based on validation performance, with a threshold of 20\% iterations without improvement.
\end{itemize}

\section{More Results of Experiment}
We evaluate the proposed framework on both the China A-share and U.S. equity markets. Our experiments are designed to: (1) demonstrate that the proposed context-free grammar provides practical advantages over linear generation methods (e.g., Reverse Polish Notation) for representing and generating alpha factors; (2) validate that the syntax representation learning method using Tree-LSTM to encode state outperforms linear network architectures; (3) evaluate the performance of the grammar-aware discovery framework across multiple metrics in comparison with existing factor-mining methods; (4) assess whether the alpha factors discovered by our model deliver superior trading performance in realistic backtesting scenarios; and (5) examine how our model enhances the performance of existing classical factors.
 
\subsection{Data}
\label{sec:data}

 For the A-share market, we adopt the constituent stocks of the CSI~300 index, and for the U.S. market, we use the constituent stocks of the S\&P~500 index. The dataset is temporally partitioned into three subsets: the training set (2010-01-01 to 2017-12-31), the validation set (2018-01-01 to 2019-12-31), and the testing set (2021-01-01 to 2024-12-31). To avoid distortions caused by abnormal market volatility and structural irregularities during the COVID-19 pandemic, data from calendar year 2020 are excluded by design. Six raw stock-level features are used as model inputs: $\{\mathrm{open},\;\mathrm{close},\;\mathrm{high},\;\mathrm{low},\;\mathrm{volume},\;\mathrm{vwap}\}$. Formulaic alpha factors are constructed by applying arithmetic operators to these base features under the grammar constraints described earlier. The prediction target for factors is the 20-day forward return, computed using closing prices for both buying and selling, i.e., $R_t^{(20)}=\frac{\mathrm{Ref}(\mathrm{close},\,-20)}{\mathrm{close}} - 1$.

\subsection{Comparison Methods}  
\label{sec:Comparison Methods}
We evaluate three variants of grammar-constrained factor discovery method: (i) \textbf{$\alpha$-Syn} (generation constrained solely by syntactic rules) (ii) \textbf{$\alpha$-Sem} (generation constrained by both syntactic and semantic rules) (iii) \textbf{$\alpha$-Sem-$k$} (generation further restricted by a length-bounding mechanism in \Cref{alg:k-bounded-derivation}). To further validate the grammar effectiveness, we also incorporate Reverse Polish Notation (RPN).
(Specifically for \textbf{$\alpha$-Syn}, we constrain the rolling window size to be an integer constant in $\alpha$-Syn to facilitate smooth training.)

For a broader performance assessment of the entire framework, we compare our method against two state-of-the-art factor mining baselines: AlphaGen~\citep{yu2023generating} and AlphaQCM~\citep{zhu2025alphaqcm}. Both employ RPN, with AlphaGen using Proximal Policy Optimization (PPO) and AlphaQCM using distributed reinforcement learning. Additionally, GPlearn \citep{zhang2020autoalpha} is included as a symbolic-regression baseline, which generates formula trees through genetic programming. All of the above factor generation methods optimize the Information Coefficient (IC) of the linear combination of factors.

To further validate our approach, we include several widely used machine learning models as additional baselines: XGBoost~\citep{wang2023xgboost}, LightGBM~\citep{bisdoulis2024assets}, LSTM~\citep{bhandari2022predicting}, ALSTM~\citep{qin2017dual}, TCN~\citep{dai2022price}, and Transformer~\citep{mozaffari2024predictive}.The hyperparameters of these models are set according to the benchmark configurations provided by Qlib~\citep{yang2020qlib}. To mitigate the impact of randomness, all models are trained and evaluated 5 times with different fixed random seeds.

\subsection{Evaluation Metrics}
\label{sec:Evaluation Metrics}
We evaluate factor effectiveness from two complementary perspectives: correlation metrics, including IC, RankIC, ICIR, and RankICIR, capture the statistical relationship between factors and future returns. 
Backtesting metrics, which are obtained by  investment simulation using a top-k/drop-n strategy (see the next paragraph for details ), including MaxDD and Sharpe, assess the profitability and risk characteristics of factors in simulated trading (see \Cref{tab:evaluation_metrics} for details).

Top-$k$/drop-$n$ strategy is applied to simulate actual trading operations: for each trading day, we first ranked stocks based on their factor prediction scores, then selected the top $k$ stocks from the sorted list. 
To balance return potential and trading costs, we adopted an equal-weight allocation approach while limiting daily portfolio adjustments to a maximum of n stocks. 
In our experiment, we set $k=60$ and $n=5$, ensuring sufficient portfolio diversification while controlling transaction costs. 

\Cref{tab:evaluation_metrics} provides the specific calculation methods for all evaluation metrics. 

\begin{table}[htbp]
  \centering
  \caption{Summary of Evaluation Metrics}
  \scalebox{0.8}{
    \begin{tabular}{>{\raggedright\arraybackslash}p{0.16\linewidth}
                    >{\raggedright\arraybackslash}p{0.16\linewidth}
                    >{\centering\arraybackslash}p{0.08\linewidth}
                    >{\raggedright\arraybackslash}p{0.28\linewidth}
                    >{\raggedright\arraybackslash}p{0.25\linewidth}}
      \toprule
      \textbf{Category} & \textbf{Metric Name} & \textbf{Abbrev.} & \textbf{Formula} & \textbf{Description} \\
      \midrule
      Correlation Metrics & Information Coefficient & IC & $\mathrm{IC} = \rho(\alpha_i, R_i)$ & Pearson correlation between factor values $\alpha_i$ and future returns $R_i$. \\
      & Rank Information Coefficient & RankIC & $\mathrm{RankIC} = \rho(r(\alpha_i), r(R_i))$ & Spearman correlation after ranking; $r(\cdot)$ is the rank function. \\
      & Information Ratio & ICIR & $\mathrm{ICIR} = \dfrac{\overline{\mathrm{IC}}}{\sigma_{\mathrm{IC}}}$ & Ratio of mean IC to its volatility, measuring prediction stability. \\
      & Rank Information Ratio & RankICIR & $\mathrm{RankICIR} = \dfrac{\overline{\mathrm{RankIC}}}{\sigma_{\mathrm{RankIC}}}$ & Ratio of mean RankIC to its volatility, evaluating rank correlation stability. \\
      \midrule
      Backtesting Metrics & Maximum Drawdown & MaxDD & $\mathrm{MaxDD} = \max_{t}\dfrac{P_{\max}(0,t) - P_t}{P_{\max}(0,t)}$ & Largest peak-to-trough decline in backtest; $P_t$ is NAV, $P_{\max}(0,t) = \max_{s \le t} P_s$. \\
      & Sharpe Ratio & Sharpe & $\mathrm{Sharpe} = \dfrac{\mathbb{E}[r_p - r_f]}{\sigma_{r_p}} \times \sqrt{N}$ & Annualized excess return per unit risk; $r_p$: daily return, $r_f$: risk-free rate, $N$: 252 (trading days). \\
      \bottomrule
    \end{tabular}
  }
  
  \label{tab:evaluation_metrics}
\end{table}

\subsection{Comparison of Different Network Architectures}
\label{sec:networks}
\begin{figure}[htbp!]
  \centering
  \includegraphics[width=0.75\linewidth]{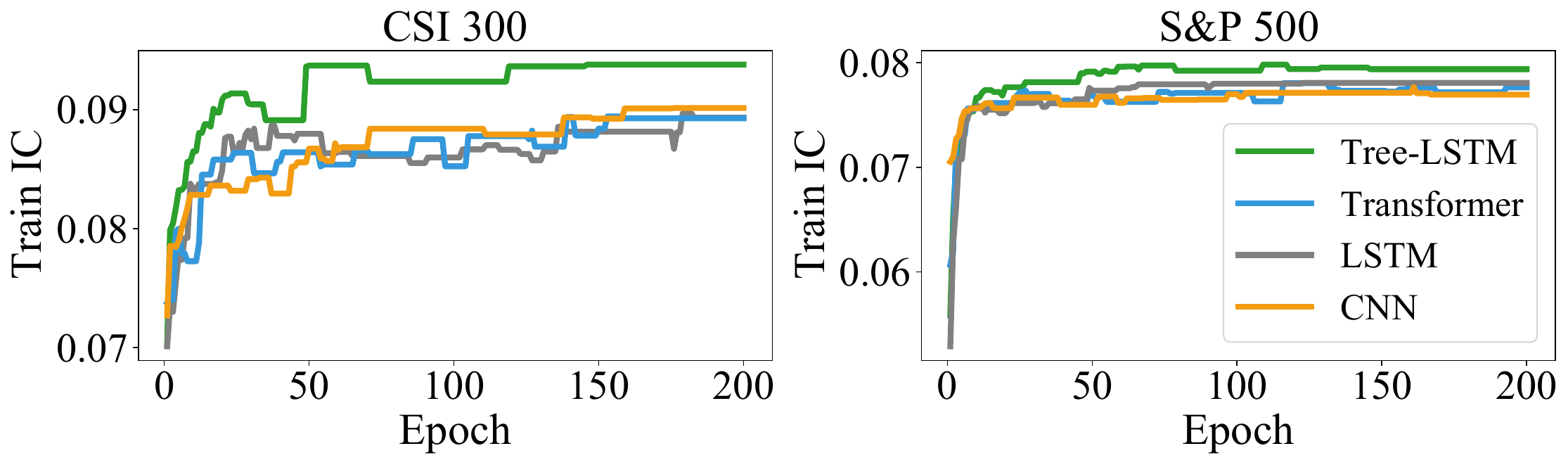}
  \caption{Comparison of training curves of different network architectures.}
  \label{fig:train_ic_2}
\end{figure}

We conducted comparative experiments under different network architectures (Transformer, LSTM, CNN) while keeping other conditions constant. With a pool size of 10 and max length 5, \Cref{fig:train_ic_2} shows training IC across epochs. Results demonstrate the effectiveness and superiority of syntax representation learning. Tree-LSTM not only extracts the structural and semantic information of expressions but also reduces redundancy caused by isomorphic forms (\Cref{def:tree_isomorphism}).

\subsection{Optimization of Combined Factor Parameters on the Validation Set}
\label{sec:valid-Parameters}

\begin{figure}[h]
  \centering
  \includegraphics[width=0.7\linewidth]{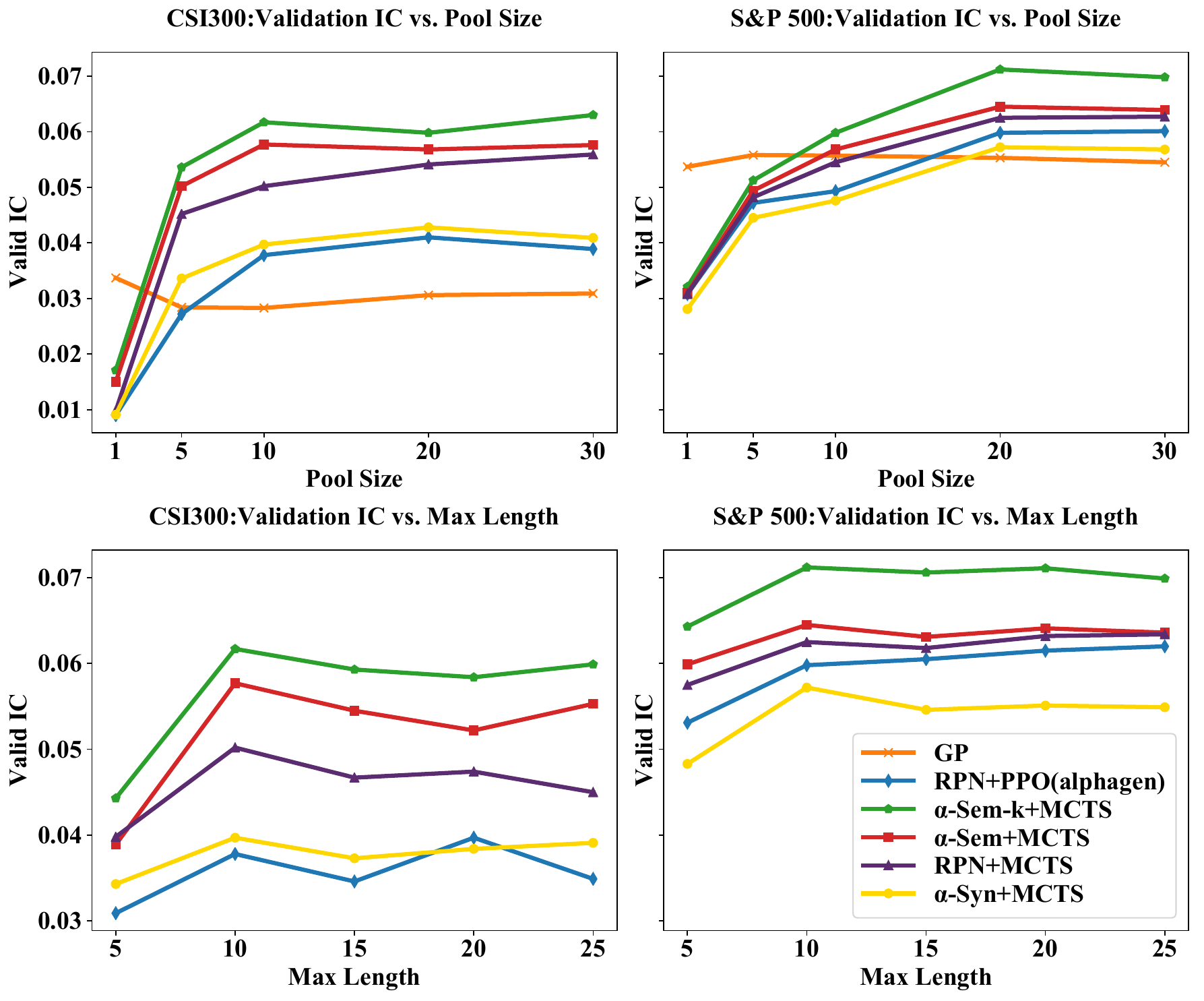}
  \caption{ Valid IC of various generation approaches.}
  \label{fig:unified_legend_ic}
\end{figure}

To obtain the optimized combined factor parameters, we conducted experiments on the validation set for two dimensions: \textit{Maximum Length of Individual Factors (Max Length)} and \textit{Factor Pool Size (Pool Size)} (results shown in \Cref{fig:unified_legend_ic}). Specifically, we first fix the maximum length of individual factors and then evaluate the valid IC for different pool sizes \{1, 5, 10, 20, 30\} to select the optimal pool size. After selecting the optimal pool size under $\alpha$-Sem-$k$, we fix it and then explore different values of the maximum length of individual factors \{5, 10, 15, 20, 25\} to identify the best configuration.

Finally, we obtain the best combined factor parameters:

\textbf{CSI~300:}
\begin{itemize}
    \item RPN+MCTS: Max Length: 10; Pool Size: 20
    \item $\alpha$-Syn: Max Length: 10; Pool Size: 20
    \item $\alpha$-Sem: Max Length: 10; Pool Size: 10
    \item $\alpha$-Sem-$k$: Max Length: 10; Pool Size: 10
    \item RPN+PPO: Max Length: 20; Pool Size: 20
\end{itemize}

\textbf{S\&P~500:}
\begin{itemize}
    \item RPN+MCTS: Max Length: 20; Pool Size: 20
    \item $\alpha$-Syn: Max Length: 10; Pool Size: 20
    \item $\alpha$-Sem: Max Length: 10; Pool Size: 20
    \item $\alpha$-Sem-$k$: Max Length: 10; Pool Size: 20
    \item RPN+PPO: Max Length: 20; Pool Size: 20
\end{itemize}

\noindent
The optimization objective of the GP method using a combined model has little effect (the generated combined factors are highly similar), so only the single-factor IC is used as its optimization objective.

\subsection{Case Study of the interpretability of formulaic factors}
\label{sec:case study}

\Cref{tab:top_alphas_final} shows an example of alpha factors generated by our framework, tested on the CSI 300 index constituents.
The mined factors exhibit strong interpretability grounded in market microstructure theory. For example, the factor \textnormal{Log(\textbar{}Std((0.05-volume),40)\textbar{})} measures the volatility of inverse trading volume over a 40-day window. This factor gauges the temporal variability of illiquidity, which may signal market stress or substantial price impact. Another example, \textnormal{Cov(volume,vwap,40)}, captures the co-movement between trading volume and the volume-weighted average price in past 40 days. A high covariance indicates strong directional consensus, potentially reflecting persistent momentum or, conversely, price reversals.

\begin{table}[htbp]
\centering
\caption{Top 10 Ranked Alphas and Their Weights}
\label{tab:top_alphas_final}
\small
\begin{adjustbox}{width=0.5\linewidth, center} 
\begin{tabular}{clr}
\toprule
\textbf{\#} & \textbf{Alpha Expression} & \textbf{Weight} \\
\midrule
1 & \textnormal{Mean(Corr(Sum(open,40),(high-volume),20),20)} & -0.00889 \\
\addlinespace
2 & \textnormal{volume} & -0.01278 \\
\addlinespace
3 & \textnormal{Std(close,40)} & 0.01778 \\
\addlinespace
4 & \textnormal{Pow(Med(Cov(high,low,30),30),0.1)} & 0.01411 \\
\addlinespace
5 & \textnormal{Delta(Log(\textbar{}Min(high,30)/0.01\textbar{}),30)} & -0.01649 \\
\addlinespace
6 & \textnormal{Cov((-0.1-Sum(close,40)),volume,20)+low} & -0.01649 \\
\addlinespace
7 & \textnormal{0.01Greater(-0.1/Corr(high,close,30),volume)} & -0.00823 \\
\addlinespace
8 & \textnormal{Log(\textbar{}Std((0.05-volume),40)\textbar{})} & 0.01224 \\
\addlinespace
9 & \textnormal{Greater(-0.01,Log(\textbar{}Log(\textbar{}low\textbar{})\textbar{}))} & -0.04616 \\
\addlinespace
10 & \textnormal{Cov(volume,vwap,40)} & -0.01412 \\
\bottomrule
\end{tabular}
\end{adjustbox}
\end{table}


\end{document}